\def\year{2020}\relax
\documentclass[letterpaper]{article} 
\usepackage{aaai20}  
\usepackage{times}  
\usepackage{helvet} 
\usepackage{courier}  
\usepackage[hyphens]{url}  
\usepackage{graphicx} 
\usepackage{graphicx}  
\usepackage{mathrsfs}
\usepackage{algorithm}
\usepackage{algorithmic}
\usepackage[noend,algo2e]{algorithm2e}
\usepackage{subcaption}
\usepackage{amsmath}
\usepackage{amssymb}
\usepackage{xcolor}
\usepackage{balance}

\newcommand{\bheading}[1]{{\noindent{\textbf{#1}}\hspace{2pt}}}

\title{Stealthy and Efficient Adversarial Attacks against Deep Reinforcement Learning}
\author{Jianwen Sun$^1$, Tianwei Zhang$^1$ , Xiaofei Xie$^{1,}$\thanks{Corresponding authors: Xiaofei Xie and Yan Zheng.}, Lei Ma$^2$, Yan Zheng$^{1,*}$, Kangjie Chen$^1$ , Yang Liu$^1$\\
$^1$Nanyang Technological University, Singapore\\
$^2$Kyushu University, Japan\\
  \{kevensun, tianwei.zhang, xfxie, yan.zheng, kangjie.chen,yangliu\}@ntu.edu.sg\\
  malei@ait.kyushu-u.ac.jp
}

\begin{document}

\maketitle

\begin{abstract}
Adversarial attacks against conventional Deep Learning (DL) systems and algorithms have been widely studied, and various 
defenses were proposed. However, the possibility and feasibility of such attacks against Deep Reinforcement Learning (DRL)
are less explored. As DRL has achieved great success in various complex 
tasks, designing effective adversarial attacks is an indispensable prerequisite towards building robust DRL algorithms.
In this paper, we introduce two novel adversarial attack techniques to \emph{stealthily} and \emph{efficiently} attack the DRL 
agents. These two techniques enable an adversary to inject adversarial samples in a minimal set of critical moments while 
causing the most severe damage to the agent. The first technique is the \emph{critical point attack}: the adversary builds a model 
to predict the future environmental states and agent's actions, assesses the damage of each possible attack strategy, and selects 
the optimal one. The second technique is the \emph{antagonist attack}: the adversary automatically learns a domain-agnostic model to 
discover the critical moments of attacking the agent in an episode. Experimental results demonstrate the effectiveness of our techniques. Specifically, to successfully attack the DRL 
agent, our critical point technique only requires 1 (TORCS) or 2 (Atari Pong and Breakout) steps, and the antagonist technique needs 
fewer than 5 steps (4 Mujoco tasks), which are significant improvements over state-of-the-art methods.

\end{abstract}

\section{Introduction}

Past years have witnessed the rapid development of Deep Reinforcement Learning (DRL) technology. The essential component of a DRL 
system is the policy, which instructs the agent to perform the optimal action responding to the environment state. Advanced
algorithms were proposed to train these policies, making them achieve significant performance on various artificial intellectual 
tasks, e.g., Go~\cite{Silver2016MasteringTG}, robotics \cite{levine2016end}, autonomous driving~\cite{sallab2017deepcar}, multiagent system~\cite{zheng2018weighted,zheng2018deep},
video game playing~\cite{Mnih2015HumanlevelCT}, testing~\cite{zheng2019wuji} and designing~\cite{zheng2019diverse}. 

A DRL policy usually adopts a Deep Neural Network (DNN) to approximate the action-value function.
However, DNNs are well known to be vulnerable to adversarial attacks~\cite{Szegedy2014IntriguingPO,fgsm}. An adversary can add
small but carefully crafted perturbations in an input, which can mislead the DNN to give an incorrect output with very high
confidence.
Extensive work have been done towards attacking supervised DNN applications across various domains such as image classification~\cite{fgsm,szegedy2015going,carlini2017towards}, audio recognization ~\cite{alzantot2018did,carlini2018audio}, and natural language processing ~\cite{alzantot2018generating}.
To date, little attention has been paid to adversarial attacks against DNN-based policies in DRL. 
As DRL is widely applied in many decision-making tasks, which require high safety and security guarantee, it is of importance
to understand the possibility and cost of adversarial attacks against DRL policies, which can help us design more robust
and secure algorithms and systems.

There are two challenges to attack a DRL system due to its unique features. First, for a supervised DNN model, the task is to
produce an output given an input. Then the adversary's goal is to make the model predict the wrong output. In contrast, the task
of a DRL is more complicated and application-specific: a self-driving car needs to reach the destination safely in the shortest time; a video game player needs to win the game 
with more points. Completion of a DRL task requires a sequence of action predictions, which are highly correlated. So the
adversary needs to compromise the agent's end goal, instead of simply misleading some predictions in a task. Second, one major
requirement for adversarial attacks is stealthiness. For the supervised DNN case, the adversary needs to add human imperceptible
perturbations to the input. For the DRL case, in addition to injecting small perturbations in one input, the adversary is also required to interfere with a very small number of steps to
cause the damage. Otherwise, crafted inputs in too many steps can be easily detected and corrected by the agent. In sum, to 
conduct a successful attack against DRL, the adversary needs to \emph{add smallest perturbations to the environment states in a minimal number of steps to incur maximum damage to the agent's end goal}.

Some attempts were made to design adversarial attacks against DRL. In \cite{huang2017adversarial}, the conventional example 
crafting technique~\cite{fgsm} was evaluated on different policy networks under both blackbox and whitebox settings. However,
it only considered the misprediction in each step, instead of the agent's end goal. Besides, it required the adversary
to attack the agent at every time step, which is less efficient and stealthy.
Two more sophisticated approaches were proposed in \cite{conf/ijcai/LinHLSLS17}. The first one is the strategically-timed attack:
the adversary dynamically decided whether the perturbation should be injected at each time step using a threshold-based 
judgment. However, this technique only considered the attack effect on one step, while ignoring the impact on the following 
states and actions. Thus the final end goal was not considered. Besides, the adversary could not control the number of steps being attacked. 
Evaluations showed that it needed 25\% of the total steps in an episode to conduct an attack, which is still too heavy. The
second technique is the enchanting attack: the adversary planned a sequence of actions for the agent to reach a damaged state. Then 
it crafted a series of adversarial examples that could lead the agent to perform the planned actions. This approach considered
the end goal and global optimization of attack steps. However, accurate prediction and enforcement of future states and actions 
are hard to achieve, especially for a long time range. Thus, this approach suffers from a low attack success rate.

In this paper, we propose two novel techniques to attack a DRL agent efficiently and stealthily. The first technique is
\textbf{Critical Point Attack}. The adversary builds a domain-specific model to predict the states of the next few steps,
as well as the damage consequences of all possible attack strategies. He adopts a Damage Awareness Metric with the criteria
of the agent's end goal to assess each attack strategy which guarantees that the adversary can select the optimal solution with a
very small number of attack steps. The second technique is \textbf{Antagonist Attack}. The adversary trains a domain-agnostic
antagonist model to identify the best attack strategy from one specific state automatically. It utilizes the agent's end goal,
the reward function as the optimization object to train this model. This model will instruct the adversary when and how to add
perturbations in order to achieve the maximum damage. 

Our attack techniques have several advantages over past work. First, we consider the global optimal 
selection of attack strategies. As such, we can significantly reduce the number of attack 
steps to achieve the same performance as past work: evaluations indicate that
Critical Point Attack only requires 1 step to attack TORCS and 2 steps to attack Atari Pong 
and Breakout, while 25\% of the total steps are required to attack the same applications in
\cite{conf/ijcai/LinHLSLS17}. Our Antagonist Attack does not need domain-specific 
knowledge, and can successfully attack Atari and Mujoco within 5 steps. Second, our techniques
are general-purpose and effective for different DRL tasks and algorithms, while the 
strategically-timed attack in \cite{conf/ijcai/LinHLSLS17} cannot work in tasks with
continuous action spaces. Third, our attacks consider long-term damage impacts. We define a
more accurate and explicit damage metric for assessment, enabling the adversary to select
the most critical moments for perturbation injection. 

The key contributions of this paper are:

\begin{itemize}
    \item design of Critical Point Attack which can discover the least number of critical moments to achieve the most severe damage.

    \item design of Antagonist Attack which can automatically discover the optimal attack strategy using the least attack cost without any domain knowledge.

    \item comprehensive evaluations on the effectiveness of the techniques against different DRL tasks and algorithms.

\end{itemize}

\section{Related Work}

\bheading{Adversarial examples against supervised DL.}
\cite{szegedy2013intriguing} found that small and undetectable perturbations in input samples could affect the results of a
target classifier. Following this initial study, many researchers designed various attack methods against image recognition applications ~\cite{fgsm,papernot2016limitations,carlini2017towards}.
Besides, adversarial examples can also be applied to speech recognition ~\cite{alzantot2018did,carlini2018audio} and natural language processing \cite{alzantot2018generating}. These techniques have been realized in physical scenarios \cite{kurakin2016adversarial,eykholt2017robust,sharif2016accessorize}.
Recently, how to systematically generate adversarial inputs were studied in DNN testing scenarios ~\cite{ma2018deepgauge,guo2019empirical,xie2019deephunter,xie2019diffchaser,du2019deepstellar}.

\bheading{Adversarial examples against DRL.}
In contrast to supervised DNN tasks, adversarial attacks against the DRL applications are still largely 
untouched. As an agent policy is usually parameterized by a DNN, the adversary can add perturbations on the 
state observations of the agent to alter his actions, which could finally affect the task's end goal.


\cite{huang2017adversarial} made an initial attempt to attack neural network policies by applying Fast Gradient 
Sign Method (FGSM)~\cite{fgsm} to the state at each time step. Follow-up work investigated the adversarial example vulnerabilities of DRL under various scenarios. Transferability of adversarial examples across different DRL models were studied in \cite{behzadan2017vulnerability}. \cite{behzadan2017whatever} evaluated the robustness and resillience of DRL against training-time and test-time attacks. \cite{pattanaik2018robust} designed methods of generating adversarial examples by maximizing the probability of taking the worst action. \cite{tretschk2018sequential} improved the training-time attack by injecting small perturbations into the training input states.
\cite{hussenot2019targeted} designed targeted attacks, which can lure an agent to consistently follow an adversary-desired policy. 
Block-box adversarial attacks against DRL were demonstrated in \cite{russo2019optimal}.

Those work proved the possibility of attacking DRL policies under different settings. However, they only focused on attacking specific states using traditional techniques, while ignoring the end goal of the entire DRL task, which consists of a continuous sequence of states. As a result, they are significantly inefficient: too many invalid and aimless perturbations cost a large amount of effort and can be easily detected by the agent.
A more interesting and challenging direction is to identify optimal moments for perturbation injection, in order to improve the attack efficiency and reduce cost.



\cite{kos2017delving} proposed some simple methods towards this goal: instead of compromising every environmental state, an adversary can compute and inject the perturbations every N frames. This strategy can reduce the injection frequency of adversarial examples compared to past work. But it still needs to attack about 120 states in each episode to affect the performance of the victim agent, as it is not smart enough to select the optimal moments. 
\cite{conf/ijcai/LinHLSLS17} proposed two advanced approaches to improve this strategy. The first one is strategically-timed attack, where an action preference function is introduced to compute the preference of the agent in taking the most preferred action over the least preferred action. The adversary will inject the perturbation if the function value is larger than a threshold. This technique considers the attack impact only on the current step but ignoring the following states. Besides, it heavily relies on the action probability distribution of the policy network output and only works for specific algorithms (A3C and DQN) and specific 
tasks (discrete action space). The second one is enchanting attack, where the adversary pre-computes a set of adversarial states and 
examples, which will lure the agent into entering the final damaged state. However, it is difficult to predict 
and control the agent's dynamic behaviors, making the approach less effective. More importantly, these 
approaches still require a relatively large number of attack steps (25\% of total steps) to succeed. 

Motivated by the above observations, in this paper we aim to further reduce the step number to achieve similar or better attack results. We propose two novel techniques to achieve this goal, as described below.

\section{Adversarial Attacks}

\subsection{Problem Statement and Attack Overview}

\begin{figure*}[htbp]
	\centering
	\includegraphics[width=\linewidth]{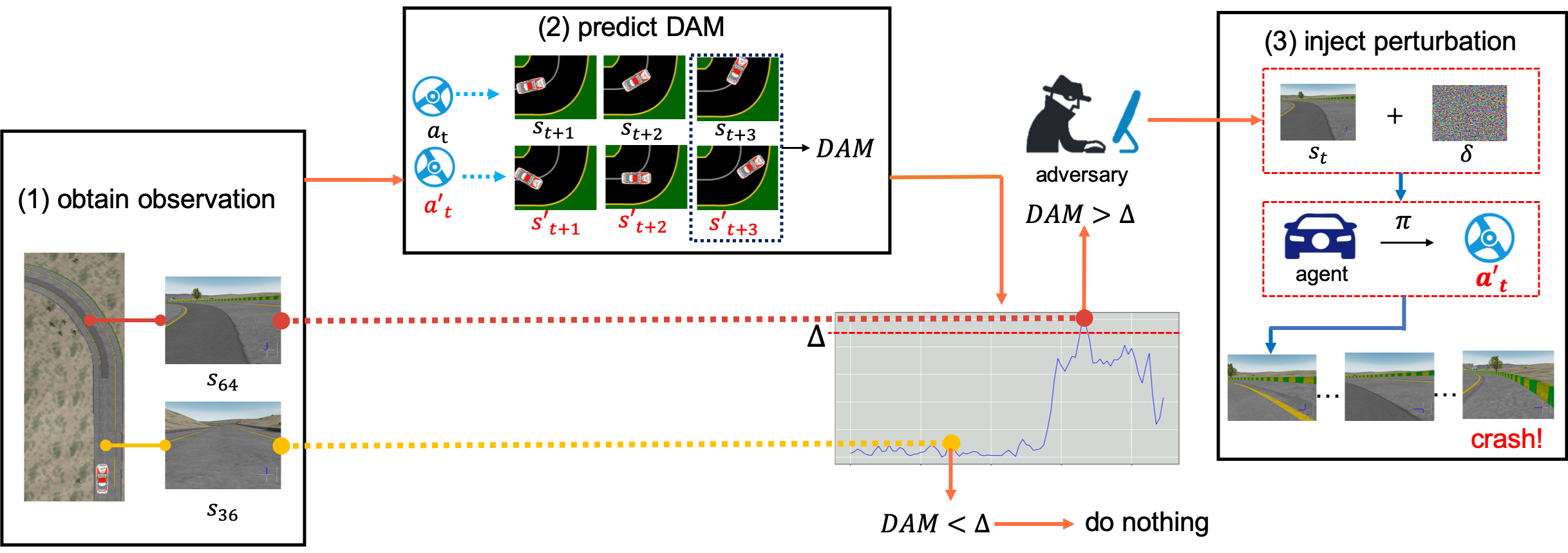}
	\caption{Illustration of Critical Point Attack on TORCS. }
    \label{methodology}
\end{figure*}

The DRL agent learns an optimal behavior through trial-and-error by interacting with an environment, which is referred to as the Markov Decision Process (MDP). Normally, the MDP can be formulated as a tuple $(\mathcal{S}, \mathcal{A}, \mathcal{P},r, \gamma)$, where $\mathcal{S}$ is the state space, $\mathcal{A}$ is the action space, $\mathcal{P}: \mathcal{S} \times \mathcal{A} \times \mathcal{S} \rightarrow [0,1]$ is the transition probability, $r(s,a) \in \mathbb{R}$ is the reward function and $\gamma \in [0,1)$ is the discount factor.
At each timestep $t$, the agent receives an observation $s_t$ and chooses an action $a_t \sim \pi(s_t)$ according to its policy $\pi$, which can be a stochastic model $\pi(s,a) \sim [0,1]$, outputting the possibility of choosing action $a$ or a deterministic policy $a = \pi(s) $ outputting the action $a$ directly. The $r(s_t,a_t)$ measures the immediate reward of $a_t$ at state $s_t$ and $R_t = \sum_{i=t...\infty} r_i $ denotes the cumulative reward. The DRL agent's goal is to learn an optimal policy $\pi^*$ that maximizes the cumulative reward $R_0$.
Without loss of generality, we consider a trajectory terminated at time $T$: $\{s_0, ... ,s_{T-1}\}$. The expected discounted reward of a policy $u$ is denoted as $R_1 = \sum_{t=0}^{T-1} \mathbb{E}_{a_t \sim u(s_t)} [ \gamma^t r(s_t, a_t) ]$.
The DRL agent aims to find a policy that can maximize $R_1$.

In contrast, the adversary in our consideration tries to decrease $R_1$ using synthesized adversarial examples, i.e., adding perturbation $\delta_t$ into the agent's observation $s_t$ and misleading the agent into performing the wrong action $a_t$.
Specifically, the adversary tries to minimize the expected reward $R_2 = \sum_{t=0}^{T-1} \mathbb{E}_{a_t \sim u(s_t+k_t\delta_t)} [ \gamma^t r(s_t, a_t) ]$, 
where the attack strategy $k_t$ denotes whether at time step $t$ the perturbation should be 
applied ($k_t = 1$) or not ($k_t = 0$). To achieve a \emph{stealthy} and \emph{efficient} 
attack, the adversary wishes to affect the agent's observations in a very small number of
steps to incur a significant decrease in $R_2$. Then the problem is: how can the adversary
find an efficient attack strategy $\{k_0, ...,k_{T-1}\}$ with the corresponding perturbations 
$\{\delta_0,...,\delta_{t-1}\}$, that can minimize $R_2$ and satisfy $\sum_{t=0}^{T-1} k_t \leq N$? Here $N$ is the upper limit of attack steps that should be far smaller than the 
episode length $T$. 


It is challenging to solve this optimization problem directly as it involves too many variables. 
Instead, we convert it into two sub-problems. The first one is ``when-to-attack'', where we will identify the optimal attack strategy $\{k_0, ...,k_{T-1}\}$. We propose two alternative approaches. The first approach is \textbf{Critical Point Attack}, which builds a model to predict the environment states and discover the critical steps for perturbations. The second technique is \textbf{Antagonist Attack}, where an adversarial agent is introduced to learn the attack strategy from the agent's policy. These two approaches can significantly reduce the number of steps required to conduct a successful attack. The second sub-problem is ``how-to-attack'', where we will figure out the corresponding perturbations $\{\delta_0,...,\delta_{t-1}\}$. We leverage existing adversarial example techniques to compute and add perturbations in the selected critical moments from the previous step. This will finally mislead the policy network to reach the adversary-desired end goal.

\subsection{Critical Point Attack}

The key idea of Critical Point Attack is inspired by human behaviors: when a human
encounters an event and has multiple actions to choose, he will always evaluate the 
consequences of each action based on his understanding of the event before making the
selection. Similarly, when the adversary encounters an environment state at one step, he 
has choices of misleading the agent to different incorrect actions, or doing nothing. The 
best strategy for him is to evaluate the attack consequences of each choice based 
on his understanding of the environment states and the agent's expected behaviors, and then select the optimal
one with the most effective results.

To establish such understanding, the adversary builds a prediction model, which can reveal 
the subsequential states and agent's actions from a starting step. At step $t$, the adversary 
considers all possible $N$ consecutive perturbations (each one can lead the agent to a different action) from this step to step $t+N-1$, predicts the environment state and evaluates the attack damage at step $t+M$ where $M \geq N$ is a predefined parameter.
The adversary picks the $N$ perturbations that can cause the most severe 
damage at step $t+M$ and adds them to the states of the following $N$ steps. Compared to the 
strategically-timed attack in \cite{conf/ijcai/LinHLSLS17}, our solution assesses and 
compares all the possible attack scenarios, and is able to discover a more optimal strategy.

Specifically, the attack is performed in two stages. First, the adversary samples the 
observations of the target environment and trains a Prediction Model ($PM$). This Prediction 
Model is a function $\mathcal{P}: (s_i, a_i) \mapsto s_{i+1}$, taking a state-action pair as 
its input and outputs a deterministic prediction of the next state. We adopt the network architecture that resembles the convolutional feedforward network from ~\cite{oh2015action}.
The input $X$ consists of state frame $s_t$ and an action $a_t$.
The loss function we use is defined as $\mathscr{L}(\theta)=\sum(s_{t+1}-\hat{s}_{t+1})^2$
where $\hat{s}_{t+1}$ is the predicted state value.
We use the 1-step prediction loss to evaluate the performance of $PM$. 
With this $PM$ and the agent's 
policy, the adversary is able to predict the subsequential states and actions from step $t$: $\{(s_t, a_t), (s_{t+1}, a_{t+1}), ...\}$. 

Second, the adversary assesses the potential damage of all possible strategies at step $t$.
We use the divergence between the post-attack state and the original one to assess the potential damage.
The intuition behind this idea is that if an attack can mislead the agent to a wrong state significantly far away from the original correct one, this can incur dangerous and severe consequence, e.g., terminating the current episode abnormally.
Specifically, a divergence function $T(s_i)$ is introduced to represent the distance indicator of the 
agent at state $s_i$.
The adversary first predicts the state $s_{t+M}$, and calculates
$T(s_{t+M})$ as the baseline. Then he considers all the possible attack strategies for the
next $N$ steps: for each step, he can use an adversarial example to make the agent perform an 
arbitrary action. For each attack strategy, he predicts the state $s'_{t+M}$, calculates
$T(s'_{t+M})$, and the Danger Awareness Metric $DAM=|T(s{'}_{t+M})- T(s_{t+M})|$. 
If this $DAM$ is larger than a threshold $\Delta$, the adversary 
will conduct the attacks in the next $N$ steps following this strategy. Otherwise, he will do 
nothing and repeat this assessment process from the next step $t+1$.

The essential point of this attack technique is to define the divergence function $T$ for one specific state. One 
possible solution is to utilize the reward function (if this is explicitly given): a smaller
reward indicates a higher possibility of termination. We empirically evaluated 
this metric, and found that the reward prediction became inaccurate when the number of 
predicted steps was larger. Instead, it is necessary to have a domain-specific definition for this
function to accurately reflect the damage of the attack on different tasks. Similar to the
domain-specific reward function, which represents the most favorable state the agent desires 
to have, divergence function $T$ should be defined as the least favorable state to reach.

Algorithm \ref{alg:cp-attack} outlines the details of Critical Point Attack.
First, the adversary calculates the baseline $s_{t+M}$ based on the prediction of $PM$, which is normally following the agent's policy.
Next, the adversary generates a set of attack strategies $A$ based on some planning algorithms.
A simple way to do this is to enumerate all possible combinations of $\{a'_t,...,a'_{t+N-1}\}$.
Then, for each attack strategy, the adversary calculates $s'_{t+M}$ with the action sequence for the 
first $N$ steps replaced by the attack.
If attack damage is large enough, we pick the $\{a'_t,...,a'_{t+N-1}\}$ to be the target sequence to make perturbations. We use a case to describe the whole attack process.

\begin{algorithm}[t]
\footnotesize
    \SetAlgoLined
    Input: $M$, $N$, $\mathcal{P}$, $T$, $s_t$\\
    Output: optimal attack strategy $\{a'_t,...,a'_{t+N-1}\}$\\
    // Calculate the baseline $s_{t+M}$ \\
    \For{$i$ \textbf{in} $\{0,...,M-1\}$}{
        $a_{t+i}=AgentAct(s_{t+i})$\\
        $s_{t+i+1}=\mathcal{P}(s_{t+i},a_{t+i})$\\
        $s_{t+i}=s_{t+i+1}$
    }
    Plan for all possible $N$-step attack strategies as a set $A$\\
    $s'_t = s_t$ \\
    \For{each attack strategy $\{a'_t,...,a'_{t+N-1}\}$ \textbf{in} $A$}{
        \For{$i$ \textbf{in} $\{0,...,M-1\}$}{
            \eIf{$i<N$}
            {
                $a_{t+i} = a'_{t+i}$ 
            }
            {
                $a_{t+i}=AgentAct(s_{t+i})$ 
            }
            $s'_{t+i+1}=\mathcal{P}(s'_{t+i},a_{t+i})$\\
            $s'_{t+i}=s'_{t+i+1}$
        }
        \If{$|T(s'_{t+M}) - T(s_{t+M})|>\Delta$}{
            return $\{a'_t,...,a'_{t+N-1}\}$
        }
    }
    return $None$
    
     \caption{Critical Point Attack at step $t$}
     \label{alg:cp-attack}
         \end{algorithm}

\bheading{Case studies.}
Fig.~\ref{methodology} illustrates an example that uses our technique to attack a 
self-driving agent. 
The adversary samples the observations of the environment in advance and trains a 
prediction model offline to predict the subsequent states in this self-driving task. At each 
step, the adversary (1) obtains observations from the environment and (2) predicts $DAM$ for each possible strate. In this case, function
$T(s_i)$ is defined as the distance between the car and the center of the road. This car will 
be more dangerous (i.e., more likely to collide) if this state value is high. The adversary 
compares each $DAM$ with $\Delta$. At step 36, all the possible strategies give smaller
$DAM$, so the adversary chooses not to conduct attacks. In contrast, at state 64, one possible
attack strategy can incur significant damage at step 67, and the corresponding $DAM$ is 
larger than $\Delta$. So the adversary will select step 64 as the critical point and (3) inject perturbation. 

\subsection{Antagonist Attack}
Critical Point Attack is very effective and it can reduce the number of required attack steps to 1 or 2 as evaluated in a later section.
However, it requires the adversary to have the domain knowledge to define the divergence 
function. To relax this assumption, we propose an enhanced attack, Antagonist Attack,
which is domain-agnostic, and general enough to be applied to different tasks easily.

The key idea of Antagonist Attack is the introduction of an adversarial agent, dubbed 
antagonist. This antagonist aims to learn the optimal attack strategy automatically without
any domain knowledge. Then at each time step, the antagonist decides whether the
adversarial perturbation should be added to the state. If so, the antagonist also decides
which incorrect action the victim agent should make to achieve the best attack effects.

Specifically, the antagonist maintains a policy $u^{adv}: s_i \mapsto (p_i, a'_i)$, mapping
the current state to the attack strategy. At each time step $t$, the antagonist observes state
$s_t$, and produces the strategy $(p_t, a'_t)$. If $p_t > 0.5$, then step $t$ is chosen
as the critical point and the adversary adds the perturbation to mislead the agent to trigger
the action $a'_t$. Otherwise, the agent follows the original action $a_t$.

The challenging part in this attack is to obtain the antagonist policy $u^{adv}(s_t)$.
Algorithm \ref{alg:ant-attack} describes the procedure of training an antagonist policy. The adversary first 
initializes the parameters of this antagonist policy sampled from random distributions.
For each step in an episode, the adversary gets the attack strategy $(p_t, a'_t)$ from
the policy $u^{adv}$, and decides whether to add perturbation or not based on $p_t$. If so,
the adversary generates the adversarial sample $s'_t$ that can mislead the agent's action
from $a_t$ to $a'_t$. He performs the action $a'_t$ and receives the reward as well as the 
next state from the environment. These data will be collected to train or update the 
antagonist policy $u^{adv}(s_t)$. The adversary trains a deep neural network with the
optimization goal of Equation \ref{eq:adv-goal}, which is very similar to the agent's goal.
The difference is that the adversary's reward $r^{adv}$ is set as the negative of the agent's
reward: $r^{adv}(s_t, a'_t) = -r(s_t,a'_t)$.
\begin{equation}
    \label{eq:adv-goal}
    \sum_{t=0}^{T-1} \mathbb{E}_{(p_t, a'_t) \sim u^{adv}(s_t),a_t \sim u(s_t)} [ \gamma^t r^{adv}(s_t, a'_t) ]
\end{equation}

Note that we set an integer variable $attack\_num$ to count the number of compromised steps.
If $attack\_num$ reaches the upper bound $N$, the antagonist will be not allowed to 
attack during the current episode.
After the training, the antagonist policy can be used to guide the adversarial attacks.

\begin{algorithm}[t]
\SetAlgoLined

 Initialize the parameters $\theta^{adv}$ of policy $u^{adv}$ \\
 \For{each episode}{

    $attack\_num = 0$ \\
    \For{each step $t$ \textbf{\emph{and}} current episode not terminated}{
    	$(p_t, a'_t) = u^{adv}(s_t)$ \\
        \eIf{$p_t>0.5$ \textbf{\emph{and}} $attack\_num < N$ }
        {
        	$s'_t = GenerateAdversarialSample(s_t, a'_t)$
            $attack\_num += 1$
        }
        {
            $s'_t = s_t$
        }
		$a''_t = u(s'_t)$ \\
        Perform $a''_t$ and receive $r^{adv}_t$ and $s_{t+1}$ \\
        Collect $(s_t, p_t, a'_t, r^{adv}_t, s_{t+1})$
    }
    Train policy $u^{adv}$ to update $\theta^{adv}$ \\
 }
 \caption{Training Antagonist Policy}
 \label{alg:ant-attack}
\end{algorithm}

\subsection{Adding Perturbation}
We can utilize either Critical Point Attack or Antagonist Attack introduced above to identify the critical moments for adversarial example injection. The next step is to generate such examples. 
Let function $F(x)=y$ be a neural network that maps input $x \in \mathbb{R}$ to output $y \in \mathbb{R}$.
In the case of m-class classifier, the output of the network is usually computed by a softmax activation function such that $y$ satisfies $0 \leq y_i \leq 1$ and $\sum_{i=0}^m y_i= 1$.
Thus the output $y$ represents a probability distribution over the class labels.
The predicted label of $x$ is $L(x) = argmax_i F(x)_i$.
The goal of an adversarial attack for a victim classifier is to find $\delta$ such that $L(x+\delta) \neq L(x)$.
If the classifier can be misled to a target label $z$, i.e. $L(x+\delta)=z \land L(x) \neq z$, $\delta$ is known as qualified adversarial perturbations.
In the case of a DRL system, we need to mislead the policy network $F{'}$, which outputs the action probability similar as a classifier network. Given the critical moment $t$ and the adversarial target action $a{'}_t$ from the previous step,
the goal of this step is to find $\delta_t$ at time step $t$ such that $argmax_i F'(s_t+\delta_t)_i = a{'}_t$.
We leverage C\&W attack technique to generate the adversarial perturbation.
This method can misguide the neural network of each individual input with 100\% success rate, which can increase the success rate of attacking the DRL's end goal.

\section{Experiments}
We evaluate the effectiveness of our proposed attacks in different types of DRL environments. 
In particular, we want to investigate the following research questions: 

\begin{itemize}
  \item \textbf{RQ1}: Are the proposed attacks generic and effective for various reinforcement learning tasks and algorithm?
  \item \textbf{RQ2}: How effective and efficient are our attacks compared to past work? 
\end{itemize}

\subsection{Experimental Settings and Implementations}
\label{targeted}

\bheading{Benchmarks.}
We select different types of DRL applications as the target victim: Atari games (Pong and Breakout),
autonomous driving (TORCS) and continuous robot control (Mojuco). For Atari games, the agents are trained with A3C algorithms. We adopt the same neural network architecture as the one in~\cite{mnih2016asynchronous}, where the policy network takes 4 continuous images as input. These input images are re-sized as 80*80 and pixel values for each image are re-scaled to [0,1]. The output of the neural network is the action probability distribution.
For TORCS environment, the agent is trained by Deep Deterministic Policy Gradient method
(DDPG)~\cite{ddpg}. The action space of TORCS is a 1-dimensional continuous steering angle, 
from -1.0 to 1.0. This environment provides sensory data of a racing car including speed, 
distance from the center of the road, engine speed, LIDAR sensor, etc.
For Mojuco tasks, the agent is trained using Proximal Policy Optimization (PPO)~\cite{ppo} as 
the policy optimizer.
All testing results of the attack methods are obtained as an average of 20 random seeds.

\bheading{Training prediction models in Critical Point Attack.}
The training data for $PM$ is sampled from the training process of the agent. In Atari games, as the image input cannot directly provide any metrics information (the position of the ball in Breakout), we turn to predict RAM data of Atari games, which provide memory information of the game and are equivalent with screen data in representing the state of Atari games. In TORCS, the observation is the sensory data that can be directly used to train $PM$. The sizes of training data and testing data are about 20M frames and 5M, respectively. The evaluation criterion of $PM$ is the mean squared error. We use a fully-connected feedforward neural network with 4 hidden layers. The output of the network is the normalized value of predicted states.

\bheading{Training antagonist policies in Antagonist Attack.}
For Atari games, we train the antagonist policies using Proximal Policy Optimization
(PPO). For the Mujoco control missions (Inverted Pendulum, Hopper, HalfCheetah, and Walker2d), 
we train the antagonists for 8M frames.

\subsection{Results of Critical Point Attack}

\bheading{Atari games.}
We first test Critical Point (CP) Attack on two Atari games: Pong and Breakout.
The essential components of Critical Point attack are the prediction model $PM$ and the
divergence function $T$. 
In Breakout,  we define $T(s) = p(s)*\delta(s)$, where $\delta(s)$ is the distance between the paddle 
and the ball, and $p(s)$ is the probability of the ball falling to the bottom (Breakout) or 
approaching the right edge (Pong).
By increasing the attack step $N$, we find that we need at least a 2-step attack to prevent the paddle from catching the ball.
As the attack effect can be captured immediately in both Pong and Breakout, we set $M$ as the number of attack steps.

\begin{figure}[t]
  \centering
  
  \begin{subfigure}{0.48\linewidth}
    \includegraphics[width=1\linewidth]{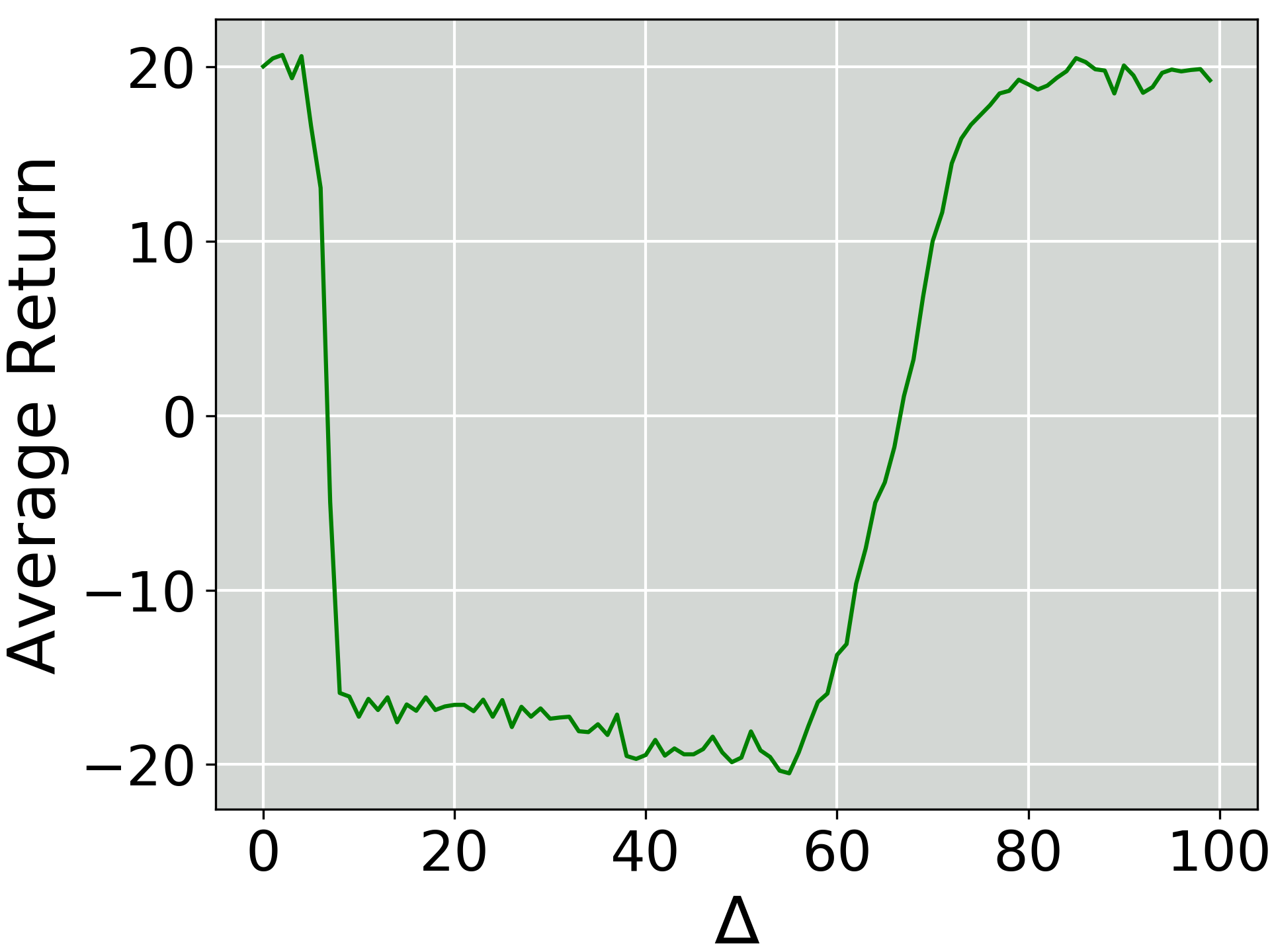}
    \caption{Pong, CP Attack }
    \label{pm-pong}
  \end{subfigure}
  \begin{subfigure}{0.48\linewidth}
    \includegraphics[width=1\linewidth]{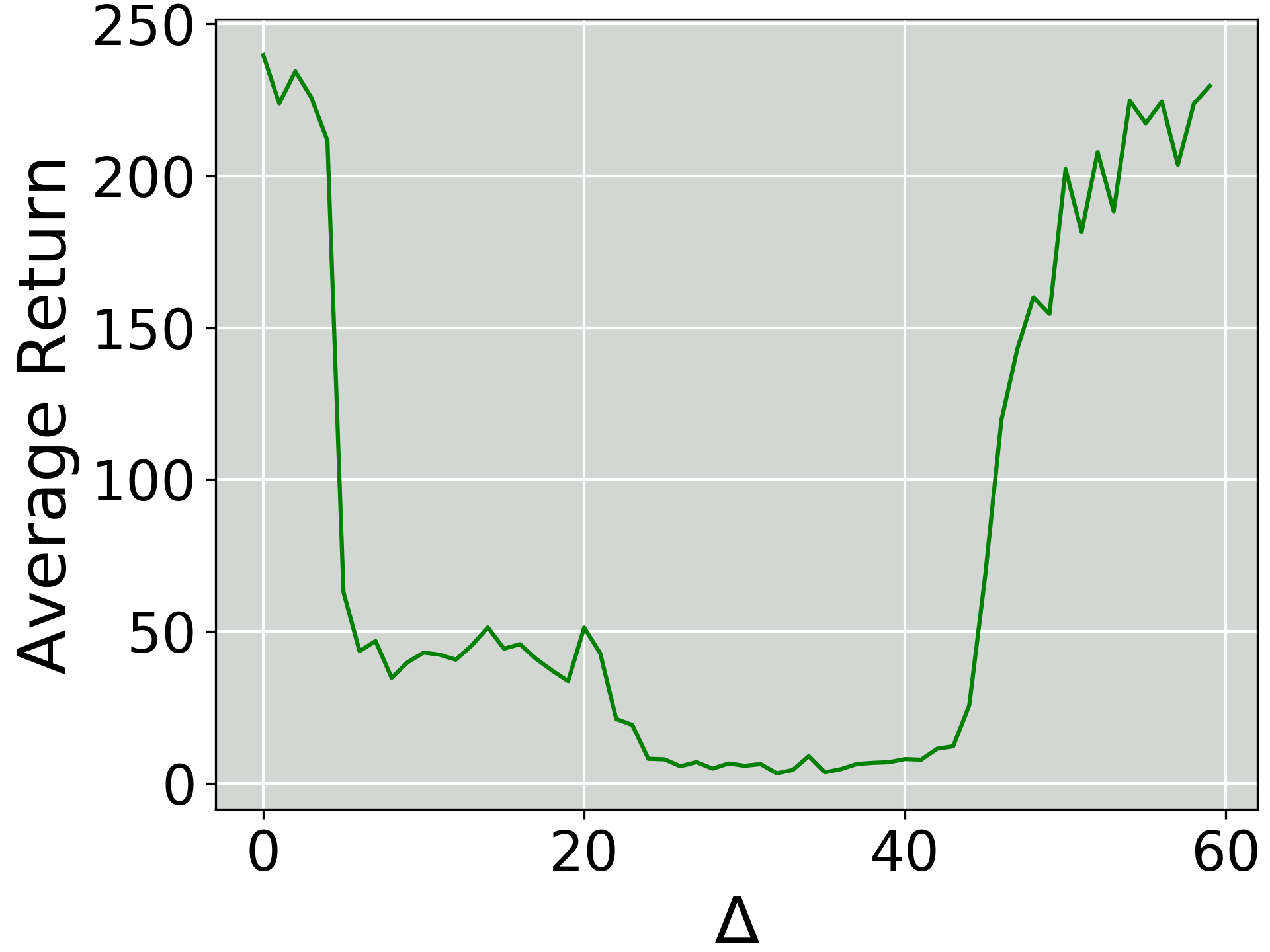}
    \caption{Breakout, CP Attack }
    \label{pm-breakout}
  \end{subfigure}
  
  \begin{subfigure}{0.48\linewidth}
    \includegraphics[width=1\linewidth]{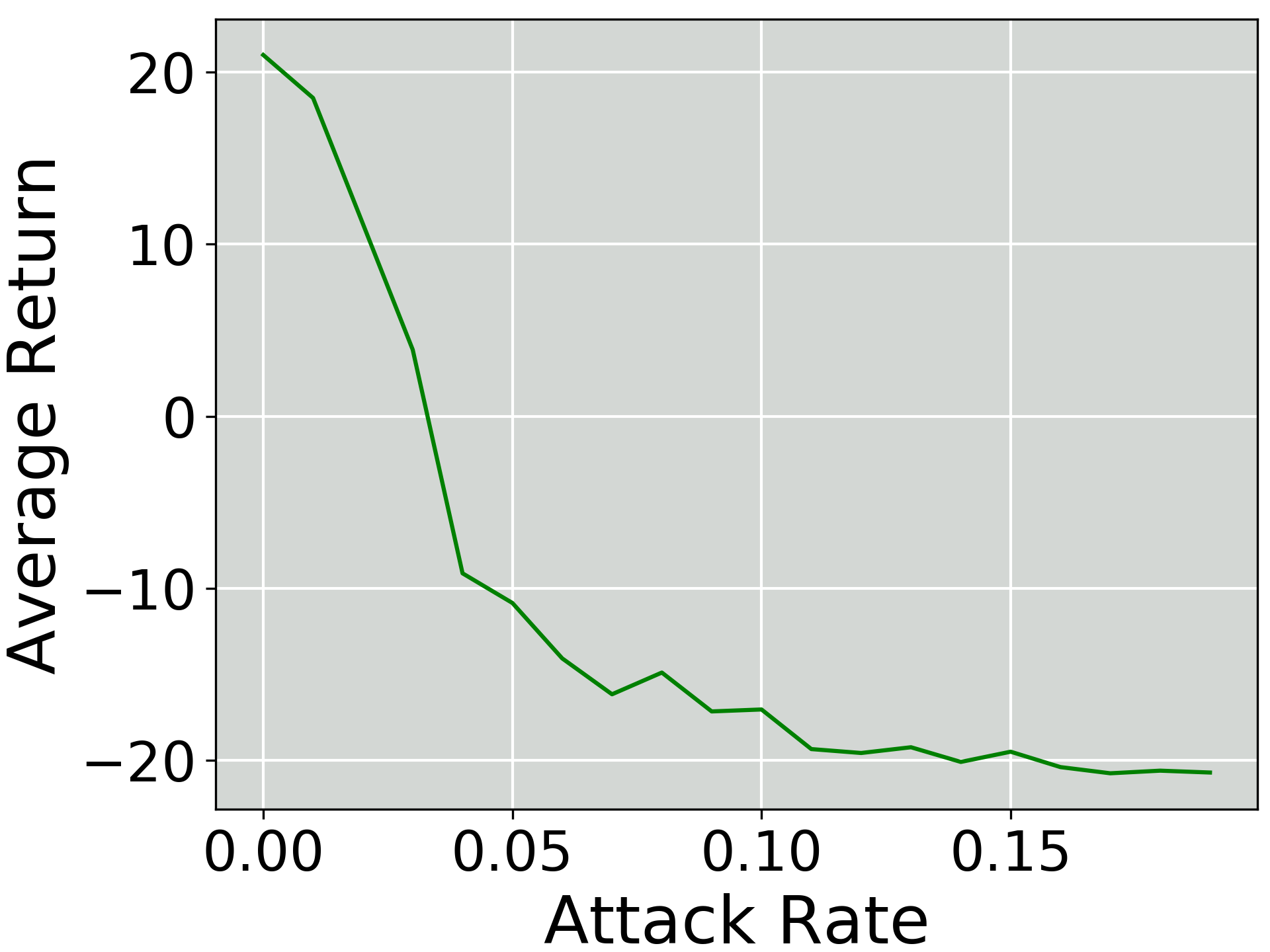}
    \caption{Pong, ST Attack }
    \label{tac-pong}
  \end{subfigure}
  \begin{subfigure}{0.48\linewidth}
    \includegraphics[width=1\linewidth]{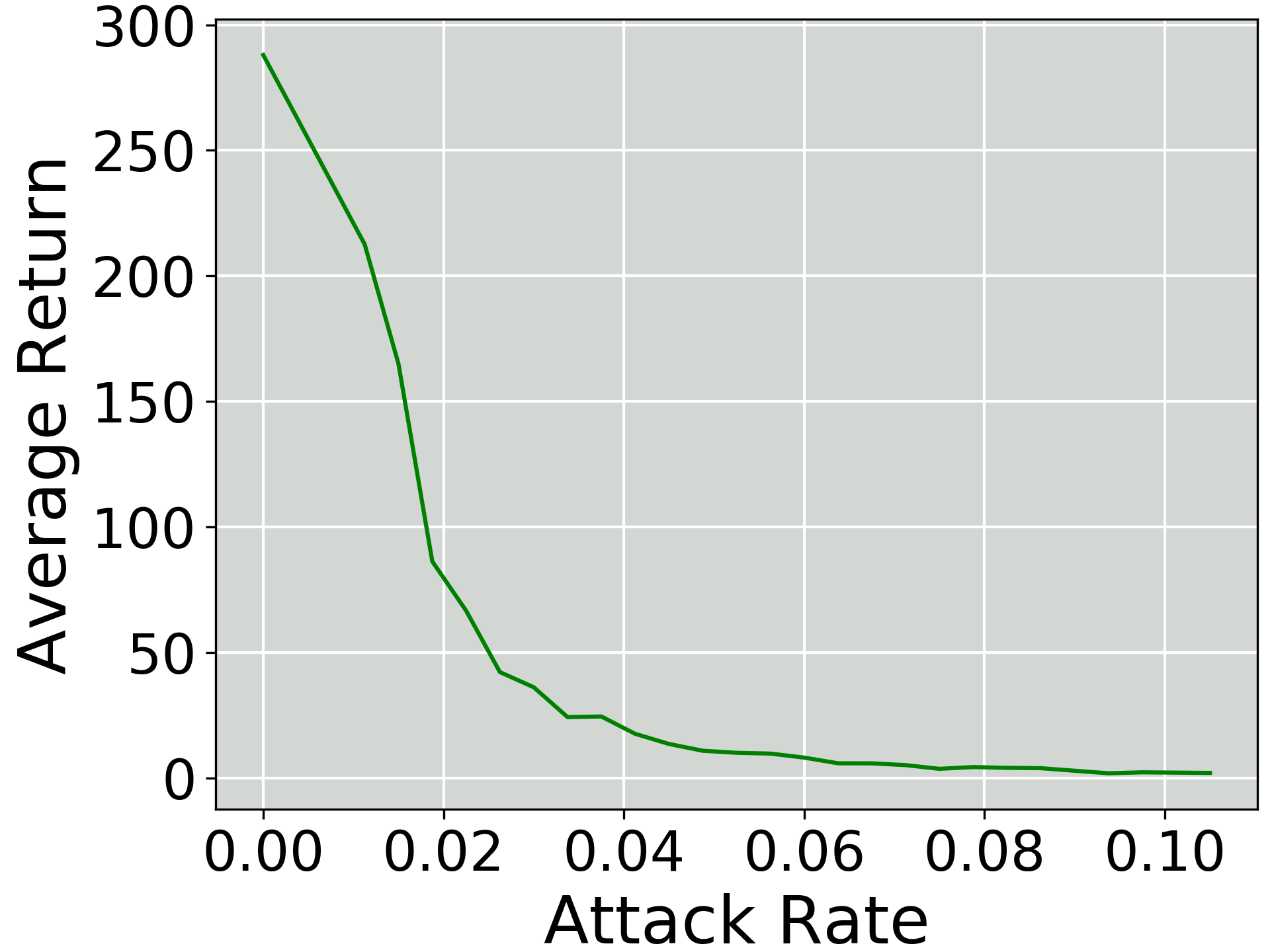}
    \caption{Breakout, ST Attack }
    \label{tac-breakout}
  \end{subfigure}
  
  \begin{subfigure}{0.48\linewidth}
    \includegraphics[width=1\linewidth]{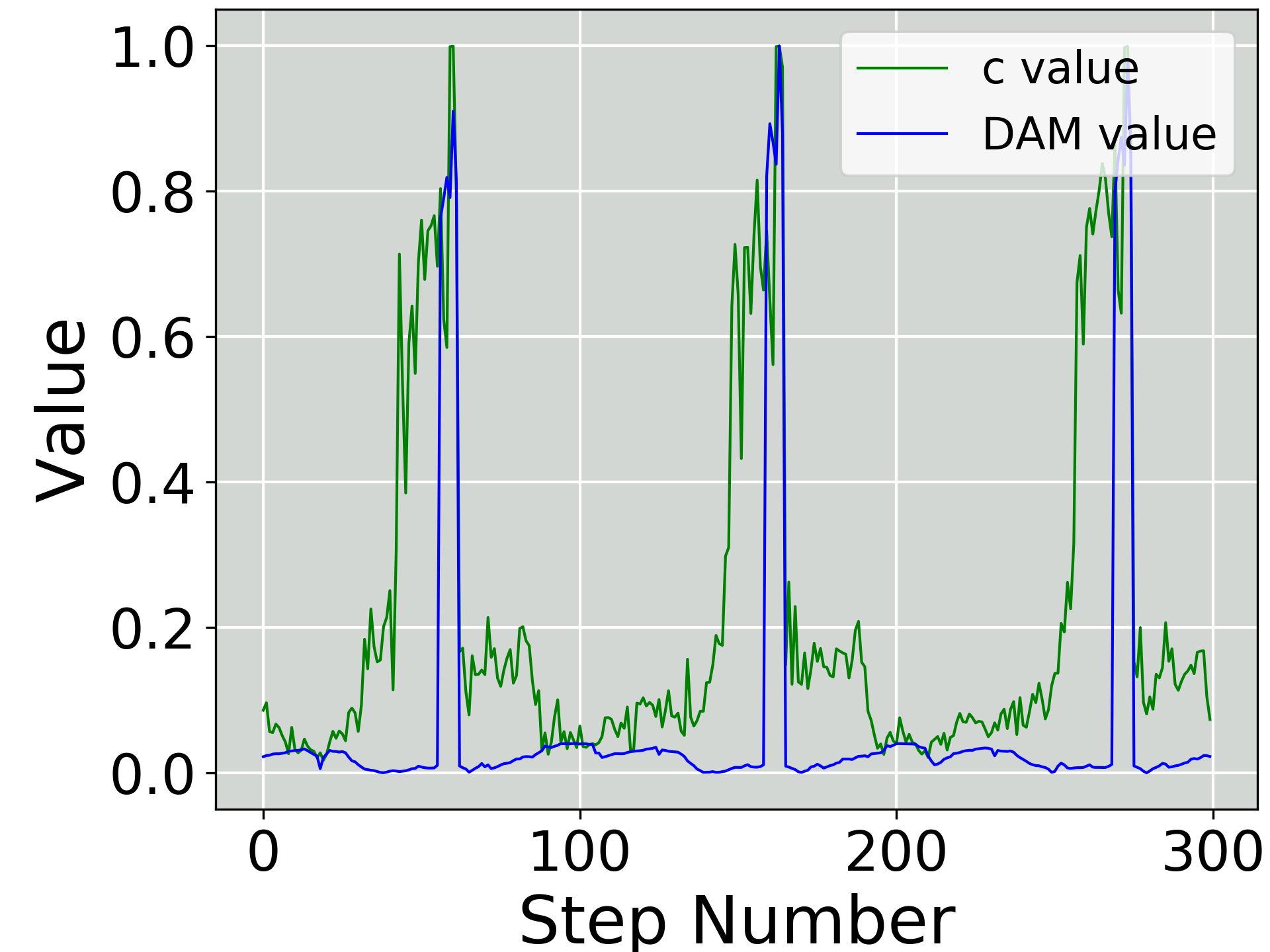}
    \caption{Pong, Attack Steps}
    \label{pm-tac-pong}
  \end{subfigure}
  \begin{subfigure}{0.48\linewidth}
    \includegraphics[width=1\linewidth]{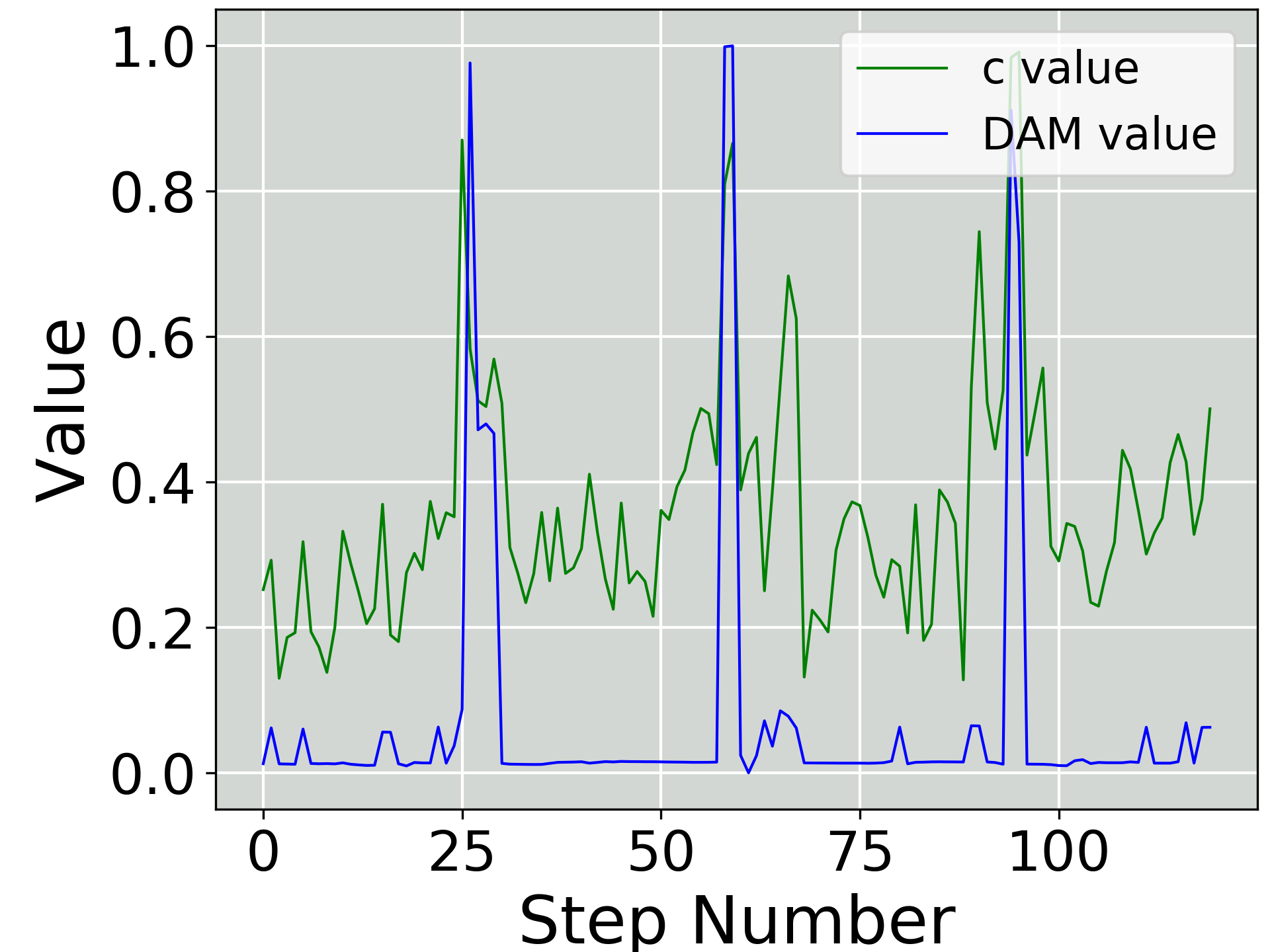}
    \caption{ {\footnotesize Breakout, Attack Steps}}
    \label{pm-tac-breakout}
  \end{subfigure}
  
  \caption{Experimental results of Atari Pong and Breakout. 
  }
  \label{atari}
\end{figure}

We launch Critical Point attack by checking $|T(s{'}_{t+2})- T(s_{t+2})|>\Delta$ with different $\Delta$ values under various initial conditions.
The results are shown in Fig.~\ref{pm-pong} and Fig.~\ref{pm-breakout}, where the y-axis denotes the accumulated reward, and the x-axis denotes different $\Delta$ values.
From the figure, we find that if we set $\Delta$ to $[9, 56]$ for Pong and $[23, 43]$ for Breakout, our method can largely reduce average return by the 2-step attack.
Lower $\Delta$ represents lower deviations from where the agent originally intended and may lead to less effective attacks.
Higher $\Delta$ improves the requirement of the dangerous value and would make it difficult to find proper timing for the attack.

We compare our approach with strategically-timed (ST) attack~\cite{conf/ijcai/LinHLSLS17}.
As shown in Fig.~\ref{tac-pong} and Fig.~\ref{tac-breakout}, ST attack 
can reach the same effect as our approach at proper attacked step rates (9.0\% for Pong and 
4.5\% for Breakout). As a comparison, the minimum attack step is 6 for Pong and 5 for 
Breakout, which is larger than the attack steps we need in CP attack.

Fig.~\ref{pm-tac-pong} and Fig.~\ref{pm-tac-breakout} compare the damage metrics of our 
approach ($DAM$, re-scaled to [0, 1]) and ST attack ($c$ value) at different 
time steps of the same trajectory. We can observe that both of the approaches give similar 
trends for selecting the critical steps. However, the metric in our approach is more precise and explicit.  

\bheading{TORCS.} The reward function in TORCS is given as:
$r_t = speed*cos(\alpha_t)-speed*trackpos$
where $speed$ is the speed of the car, $\alpha$ is the head angle, $trackpos$ is the distance 
between the car and the track axis. Based on this reward, we define $T(s) = trackpos$, as a 
larger $trackpos$ indicates that the car is in greater danger.

The parameter $N$ is set to 1 because we found that a 1-step attack is already capable of misleading the car to collide.
We set $M=3$ to assess the damage impact, as the effect of an incorrect action usually shows up after a couple of steps.
For each time step, we enumerate 200 target steering angles in [-1.0, 1.0] with the granularity of 0.01.
The adversary predicts the states after taking these steering angles based on $PM$.
The agent policy gives actions after the attack.

\begin{figure}[t]
\centering

\begin{subfigure}{0.48\linewidth}
    \includegraphics[width=1\linewidth]{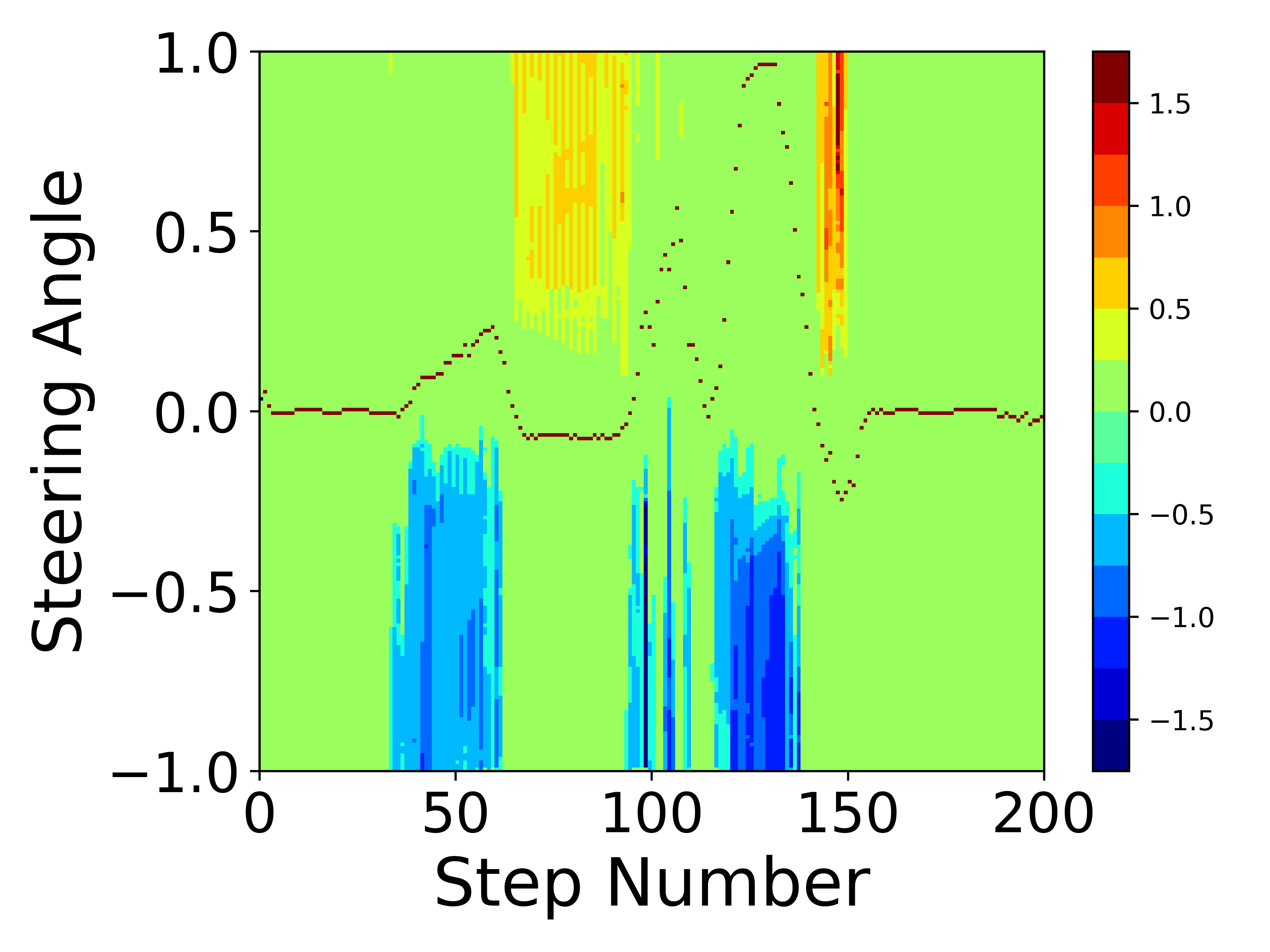}
    \caption{Predicted DAM}
    \label{torcs_pm_succ_1}
  \end{subfigure}
  \begin{subfigure}{0.48\linewidth}
    \includegraphics[width=1\linewidth]{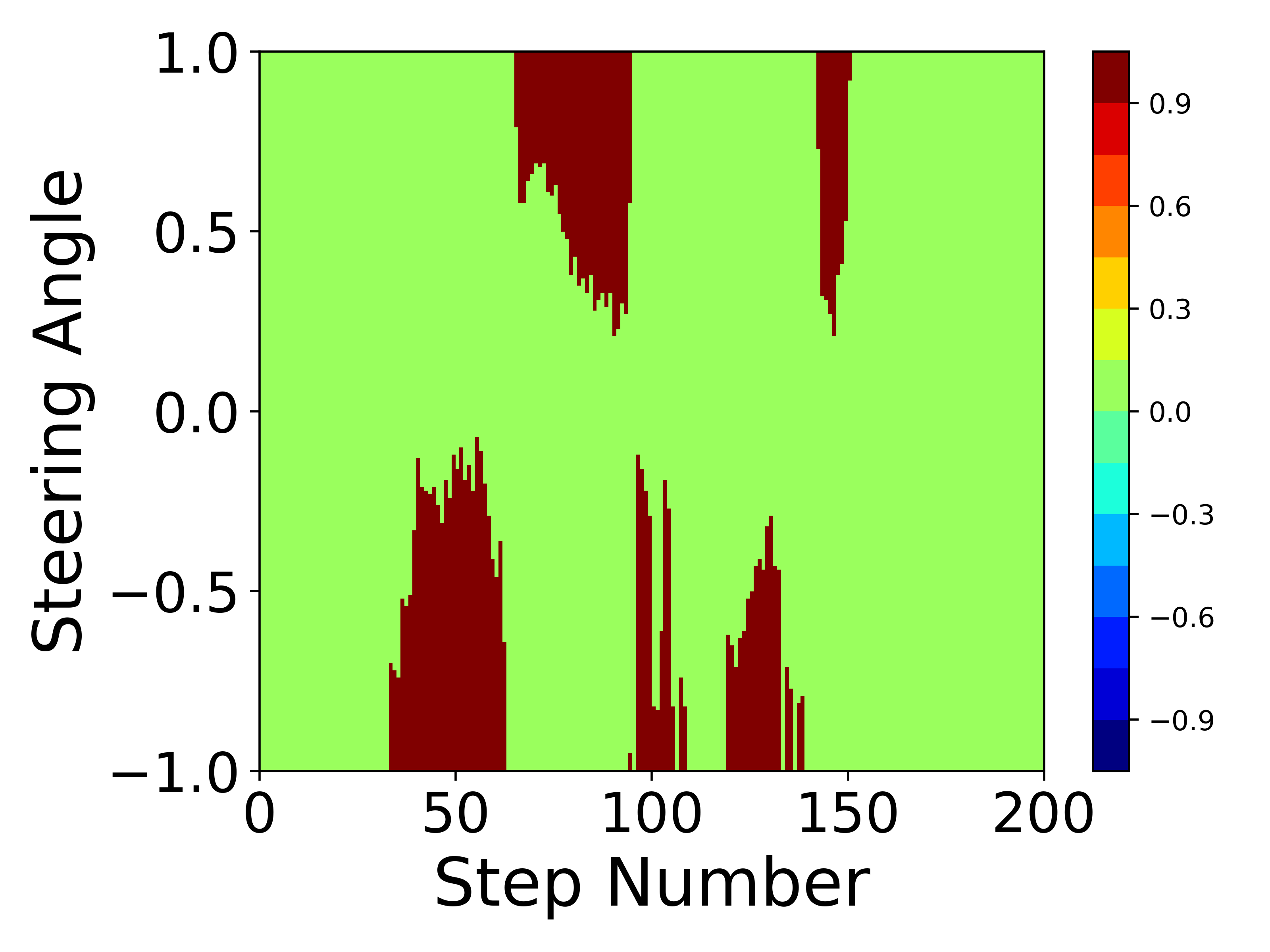}
    \caption{Crush Test}
    \label{torcs_pm_succ_2}
  \end{subfigure}

\begin{subfigure}{0.48\linewidth}
    \includegraphics[width=1\linewidth]{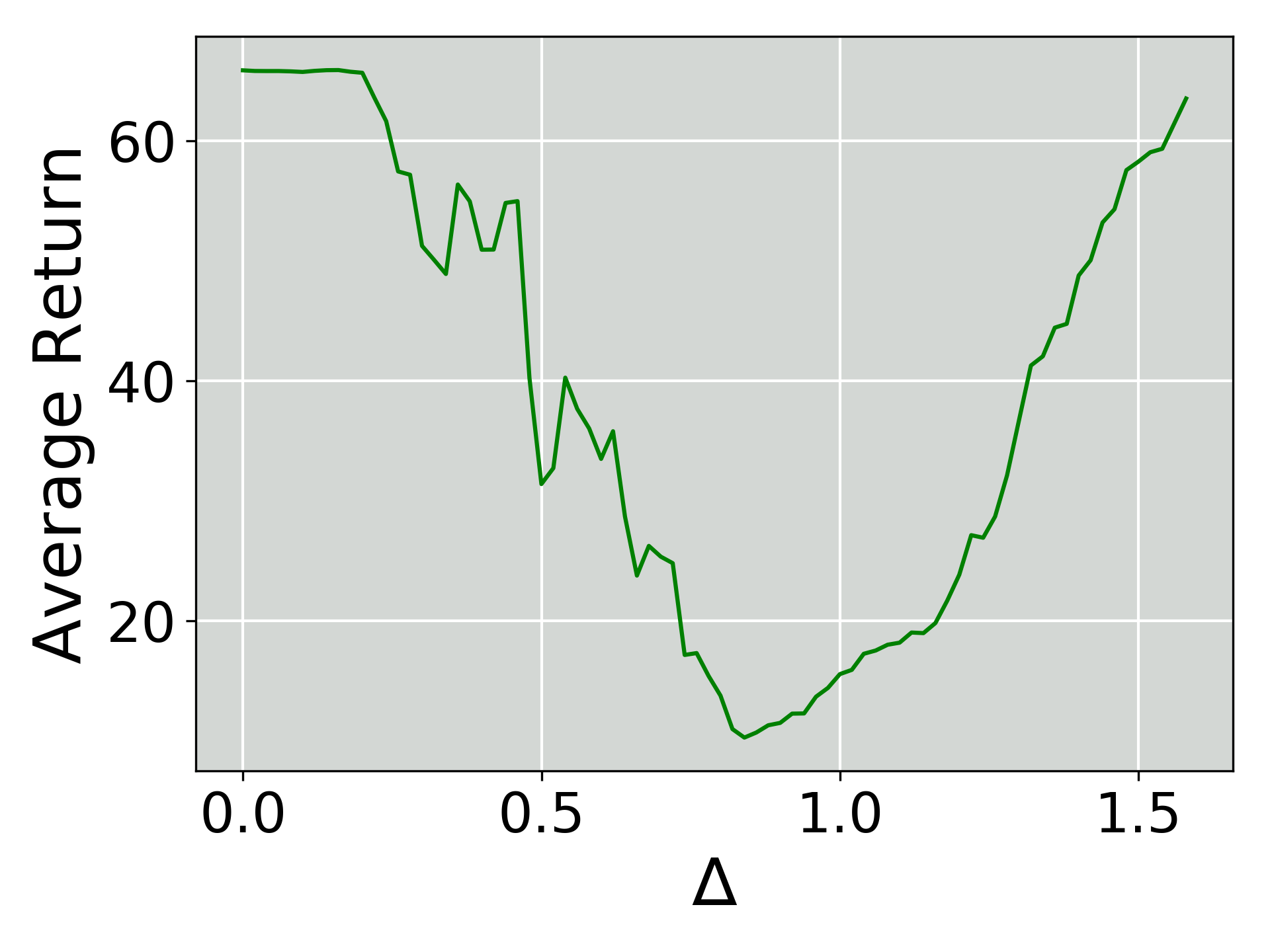}
    \caption{DDPG, CP attack}
    \label{torcs-ddpg}
  \end{subfigure}
  \begin{subfigure}{0.48\linewidth}
    \includegraphics[width=1\linewidth]{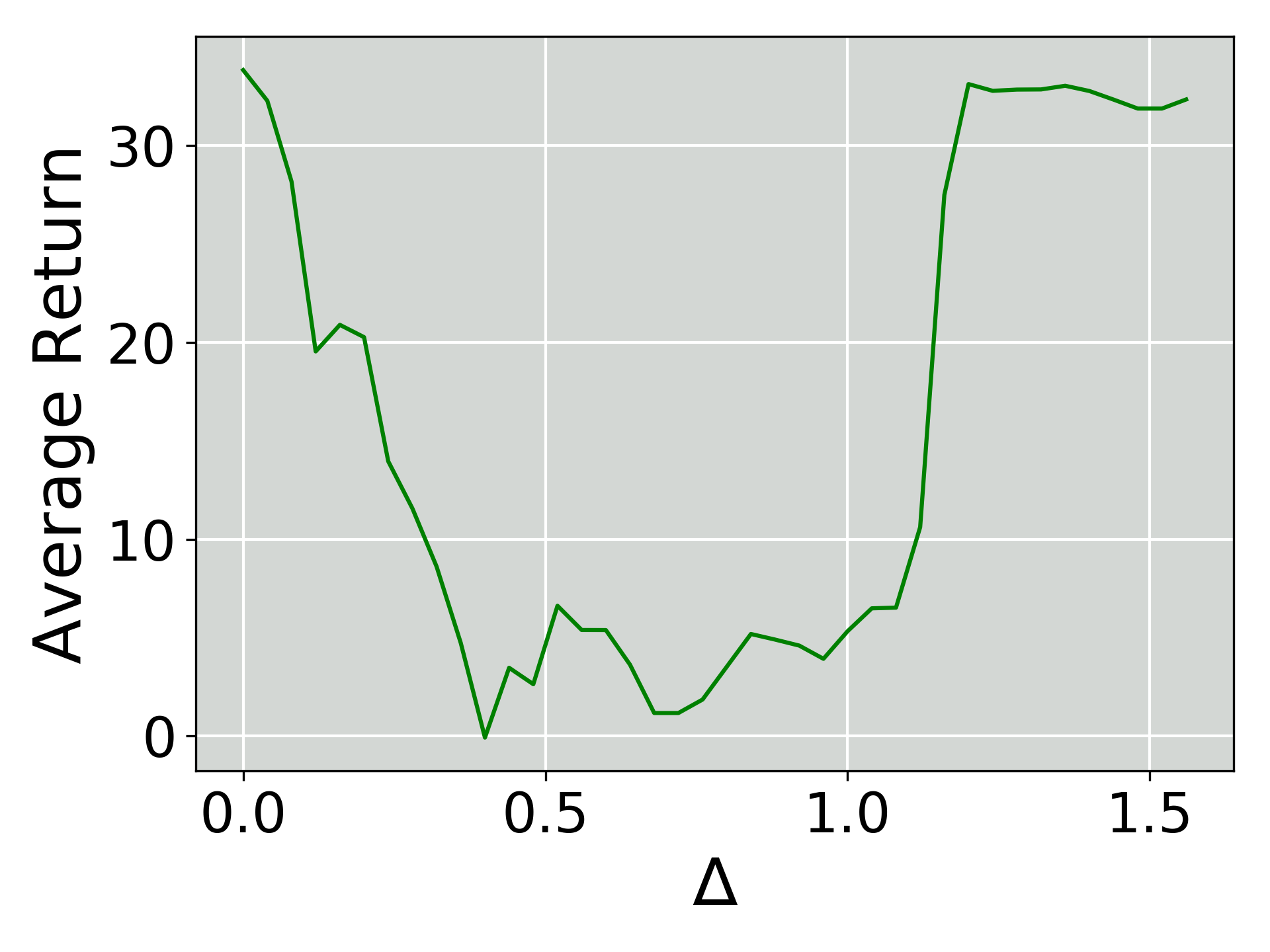}
    \caption{A3C, CP attack}
    \label{torcs-a3c-pm}
  \end{subfigure}

\begin{subfigure}{0.48\linewidth}
    \includegraphics[width=1\linewidth]{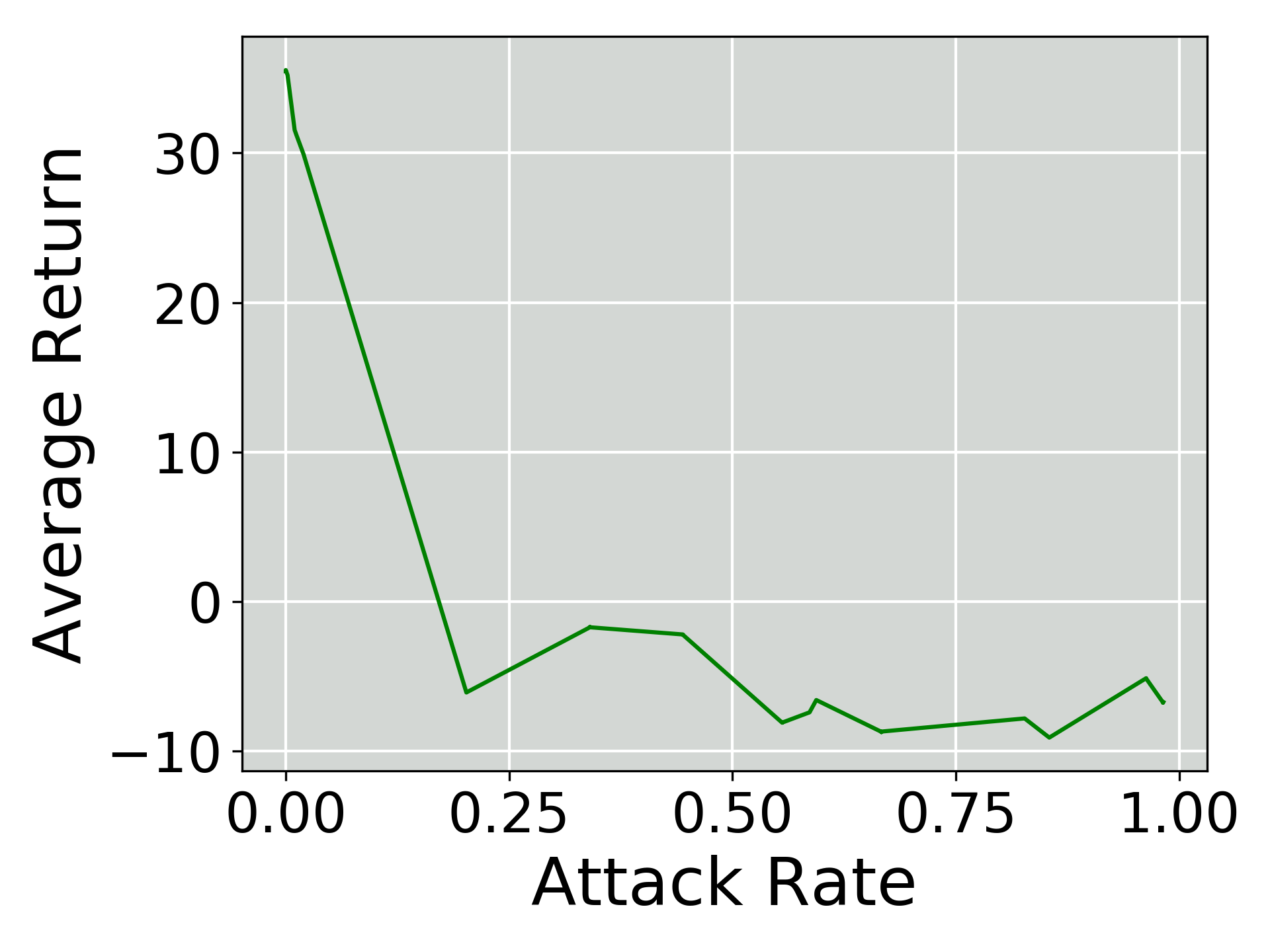}
    \caption{A3C, ST attack}
    \label{torcs-a3c-pref}
  \end{subfigure}
  \begin{subfigure}{0.48\linewidth}
    \includegraphics[width=1\linewidth]{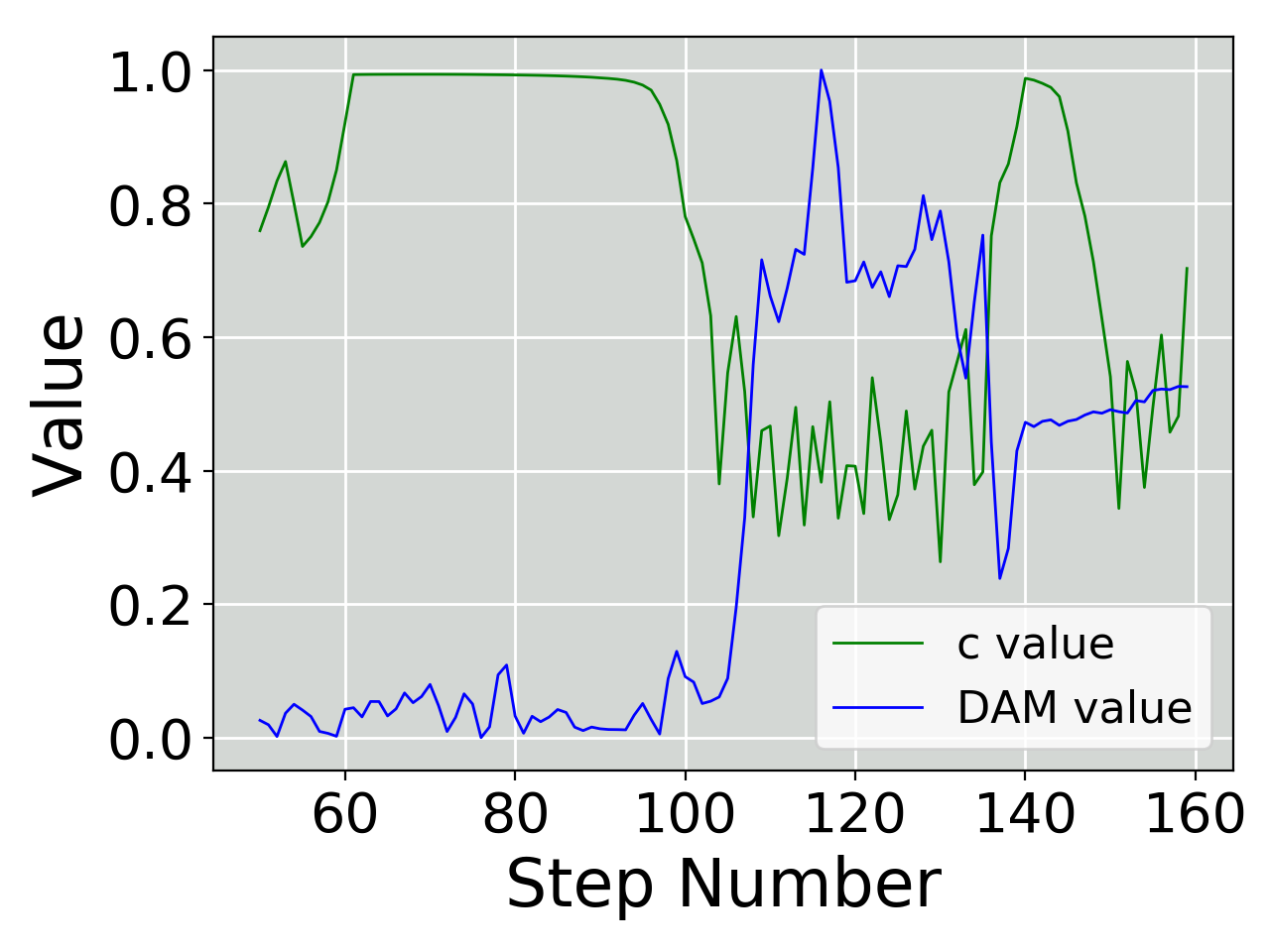}
    \caption{A3C, Attack Steps}
    \label{torcs-both}
  \end{subfigure}
  
  \caption{Experimental results of TORCS.}
  \label{torcs}
\end{figure}

In Fig.~\ref{torcs_pm_succ_1}, we sample a single trajectory of our DDPG agent and calculate $DAM$ for possible actions (y-axis) at each step (x-axis), represented as color.
The green zone means that the $DAM$ values are within the safe range ($\Delta=0.5$), while the blue zone and the red zone mean that the agent would be too close to the borders.
The dark red dots in the green zone are actions taken by the DDPG agent.
We test these attacks on the same trajectories under the same initial condition in the TORCS simulator.
The results are shown in Fig.~\ref{torcs_pm_succ_2}, where the red zone means that the agent crashes resulted from the attack, and the green zone means the agent is still safe after the attack.
Our CP attack can correctly predict the critical moments with an accuracy of 81.5\%.

We launch CP attack on the DDPG agent using different $\Delta$ values for 2,000 episodes.
Fig.~\ref{torcs-ddpg} shows the average return (y-axis) for different $\Delta$ values (x-axis).
We find that 1-step attack would be the most effective when we set $\Delta$ to about 0.82.
These attacks usually cause a car collision or loss of control.

To compare with ST attack against TORCS, we train an A3C policy in 
discrete action space. We discretize the steering wheel from -1.0 to 1.0 in 7 degrees and treat these degrees as discrete actions.
Fig.~\ref{torcs-a3c-pm} and \ref{torcs-a3c-pref} show the average return (y-axis) using our approach and ST attack for different $\delta$ and $c$ values (x-axis).
ST attack can achieve the same effect by attacking 80 steps (20$\%$ of an episode with a length of 400 time steps), which is much larger than 1 step in our approach.
The comparison results between $c$ values and $DAM$ values are shown in Fig.~\ref{torcs-both}.
As we expect, the adversary using our approach is more inclined to attack the agent when the car is close to the corner (from steps 105 to 135), because the car is unstable and it is easy for the adversary to make the car collide.
An interesting discovery is that ST attack tends to fool the agent when the car goes straight (from steps 65 to 95).
The reason is that the agent can gain more rewards at these steps. Therefore the action preference is strong even though these steps are non-critical.
This explains why ST attack takes more steps to achieve an equivalent attack goal.

\subsection{Results of Antagonist Attack}

\begin{figure}[t]
\centering

\begin{subfigure}{0.48\linewidth}
    \includegraphics[width=1\linewidth]{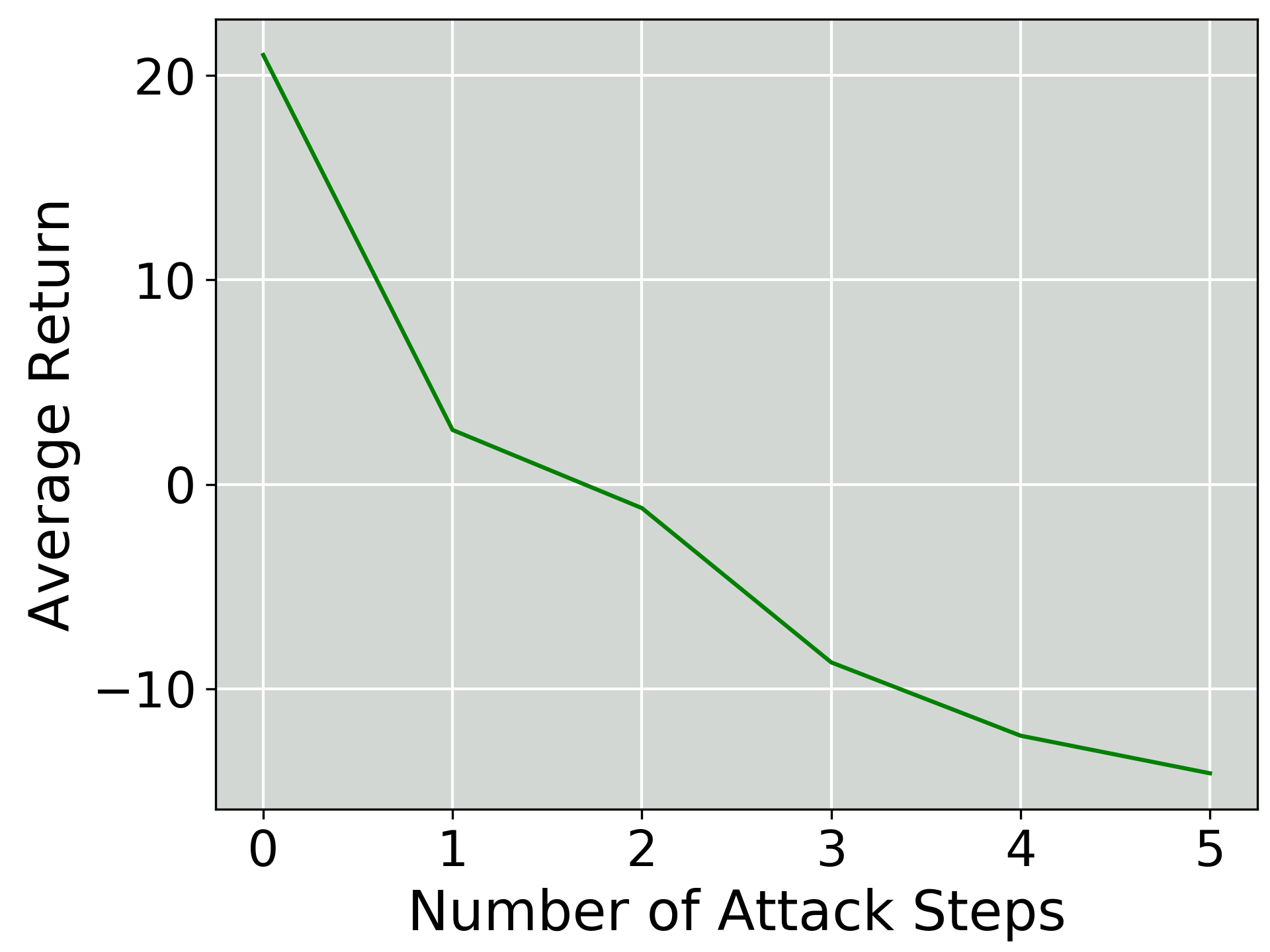}
    \caption{Pong}
    \label{anta-pong}
  \end{subfigure}
  \begin{subfigure}{0.48\linewidth}
    \includegraphics[width=1\linewidth]{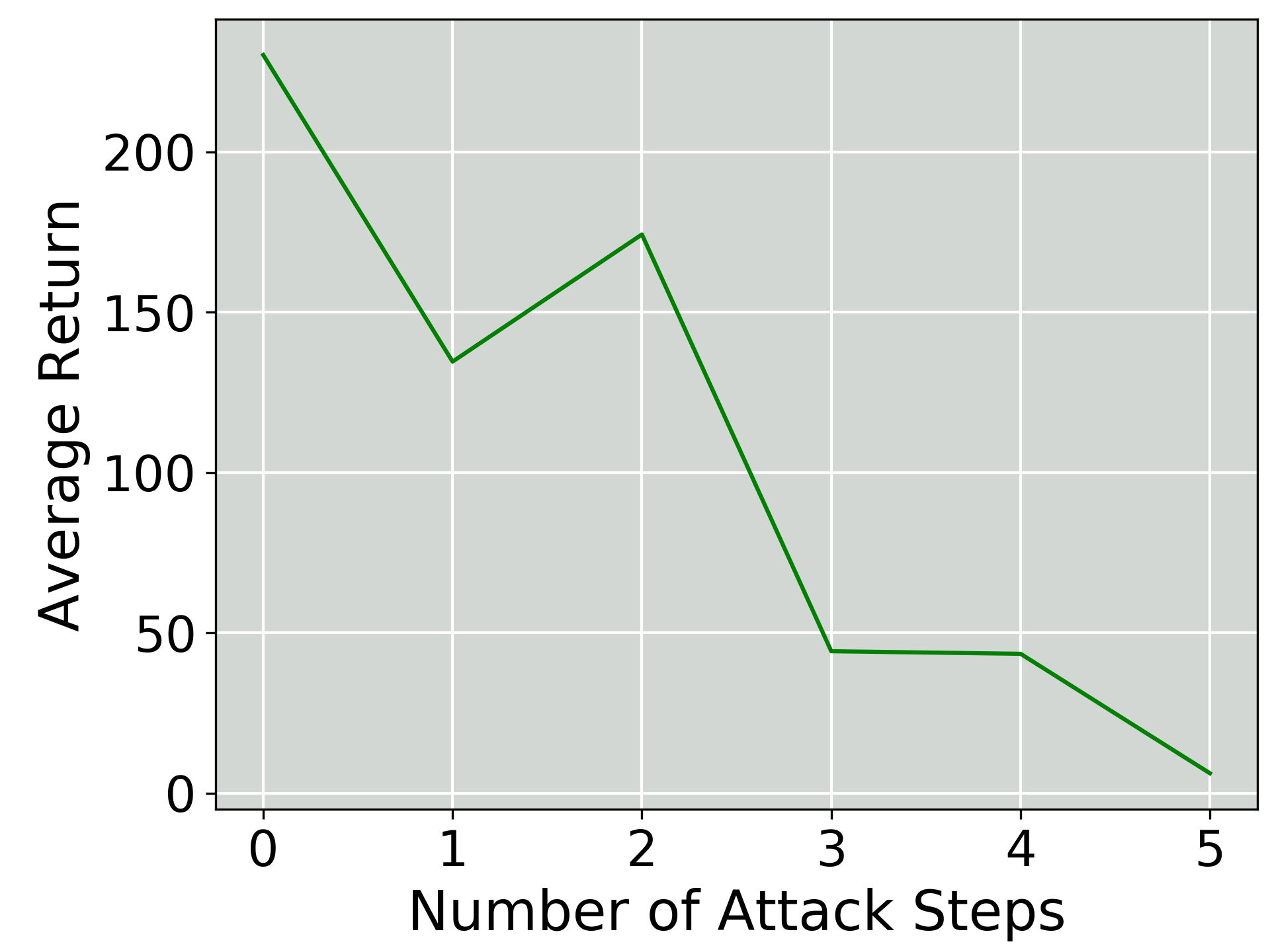}
    \caption{Breakout}
    \label{anta-breakout}
  \end{subfigure}

\begin{subfigure}{0.48\linewidth}
    \includegraphics[width=1\linewidth]{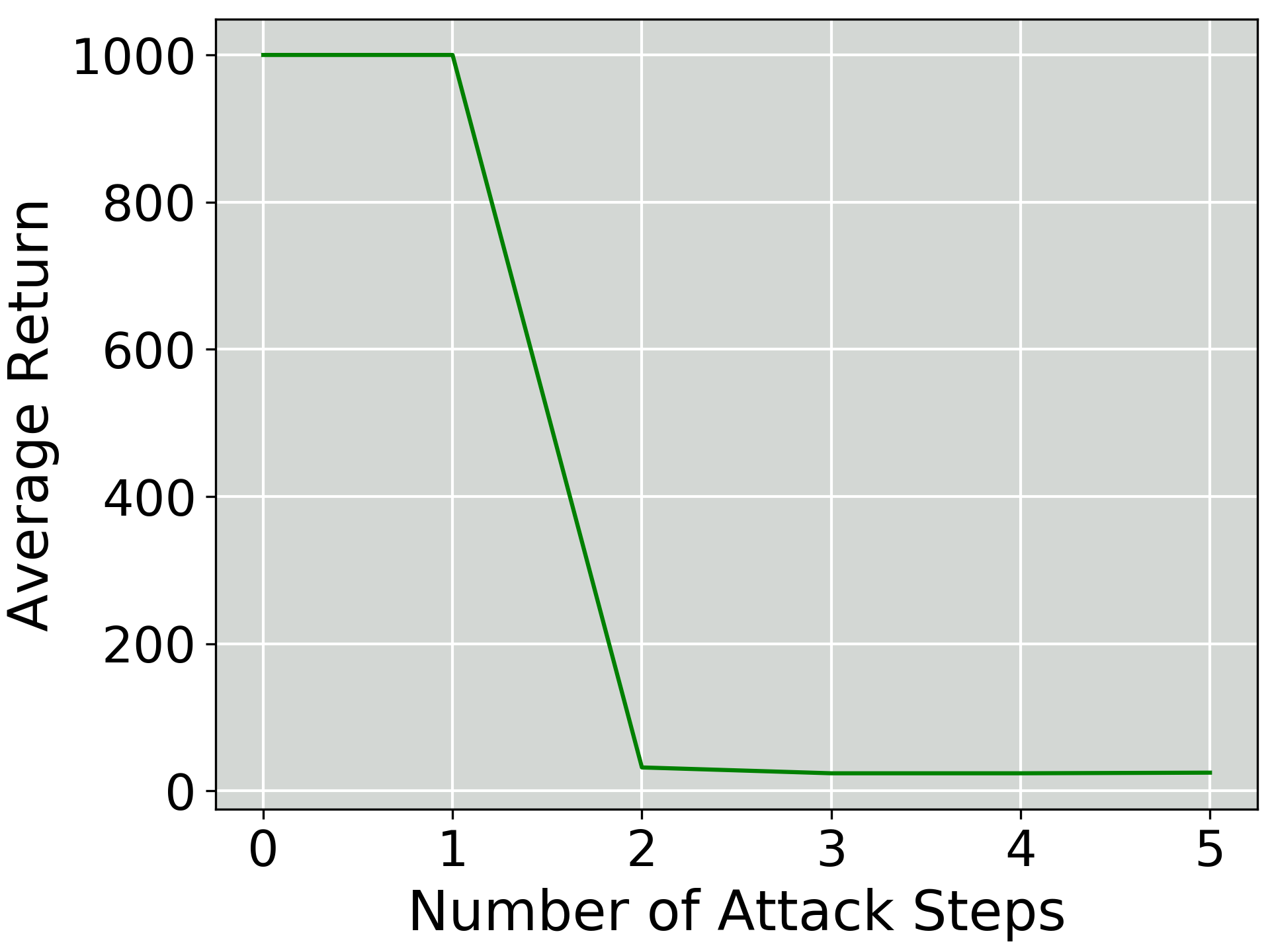}
    \caption{InvertedPendulum}
    \label{anta-inverted}
  \end{subfigure}
  \begin{subfigure}{0.48\linewidth}
    \includegraphics[width=1\linewidth]{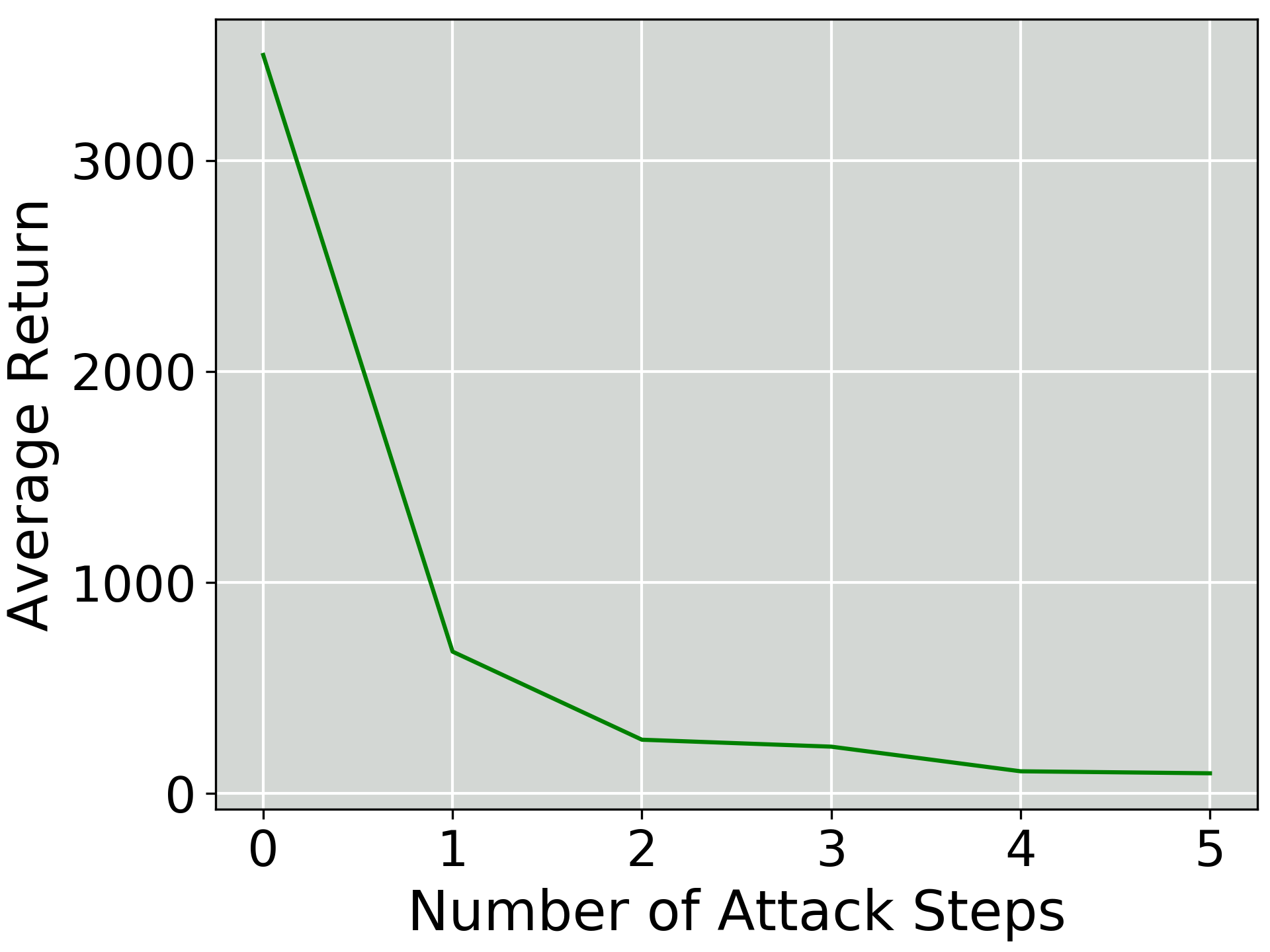}
    \caption{Hopper}
    \label{anta-hopper}
  \end{subfigure}

\begin{subfigure}{0.48\linewidth}
    \includegraphics[width=1\linewidth]{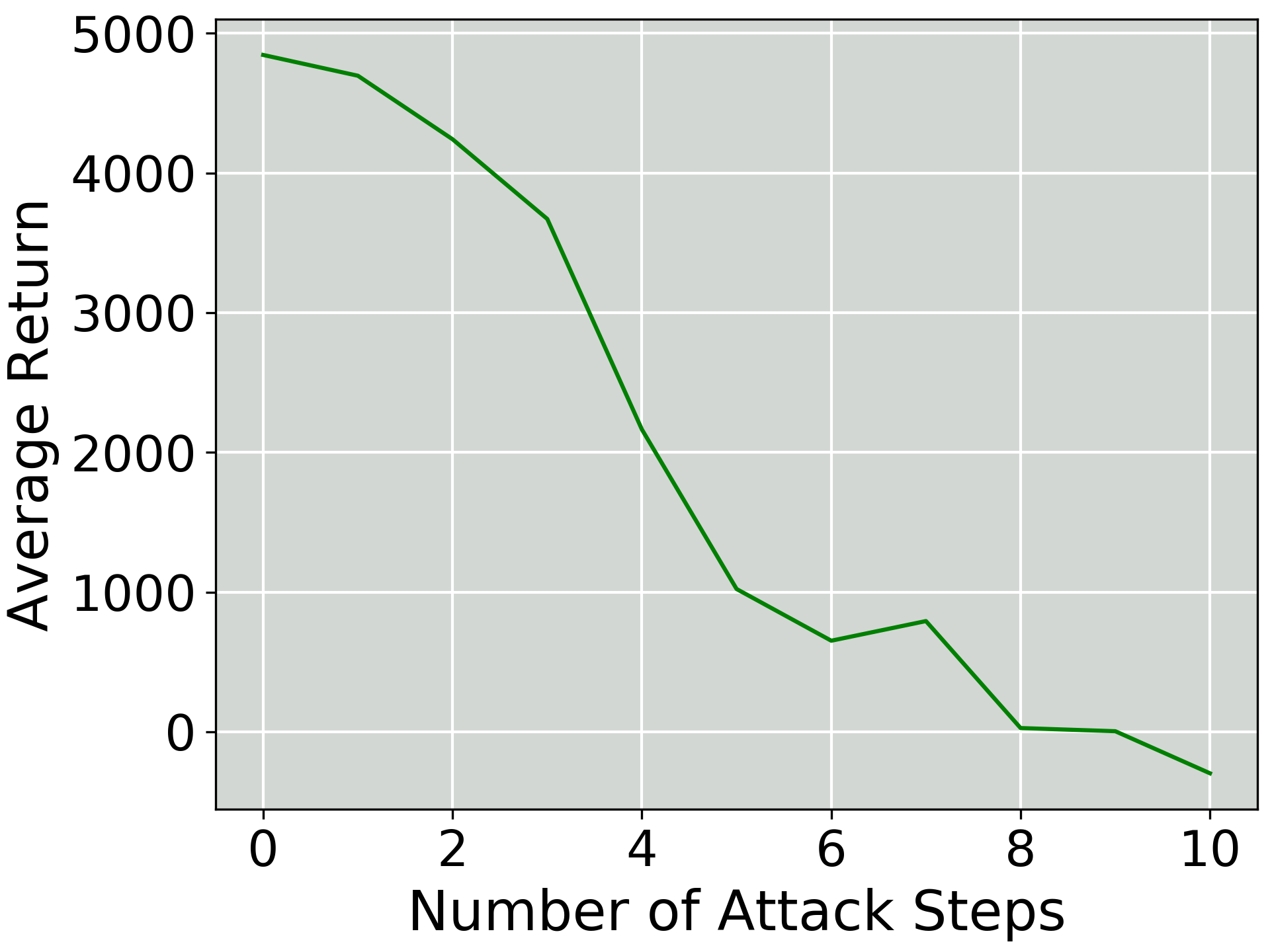}
    \caption{HalfCheetah}
    \label{anta-half}
  \end{subfigure}
  \begin{subfigure}{0.48\linewidth}
    \includegraphics[width=1\linewidth]{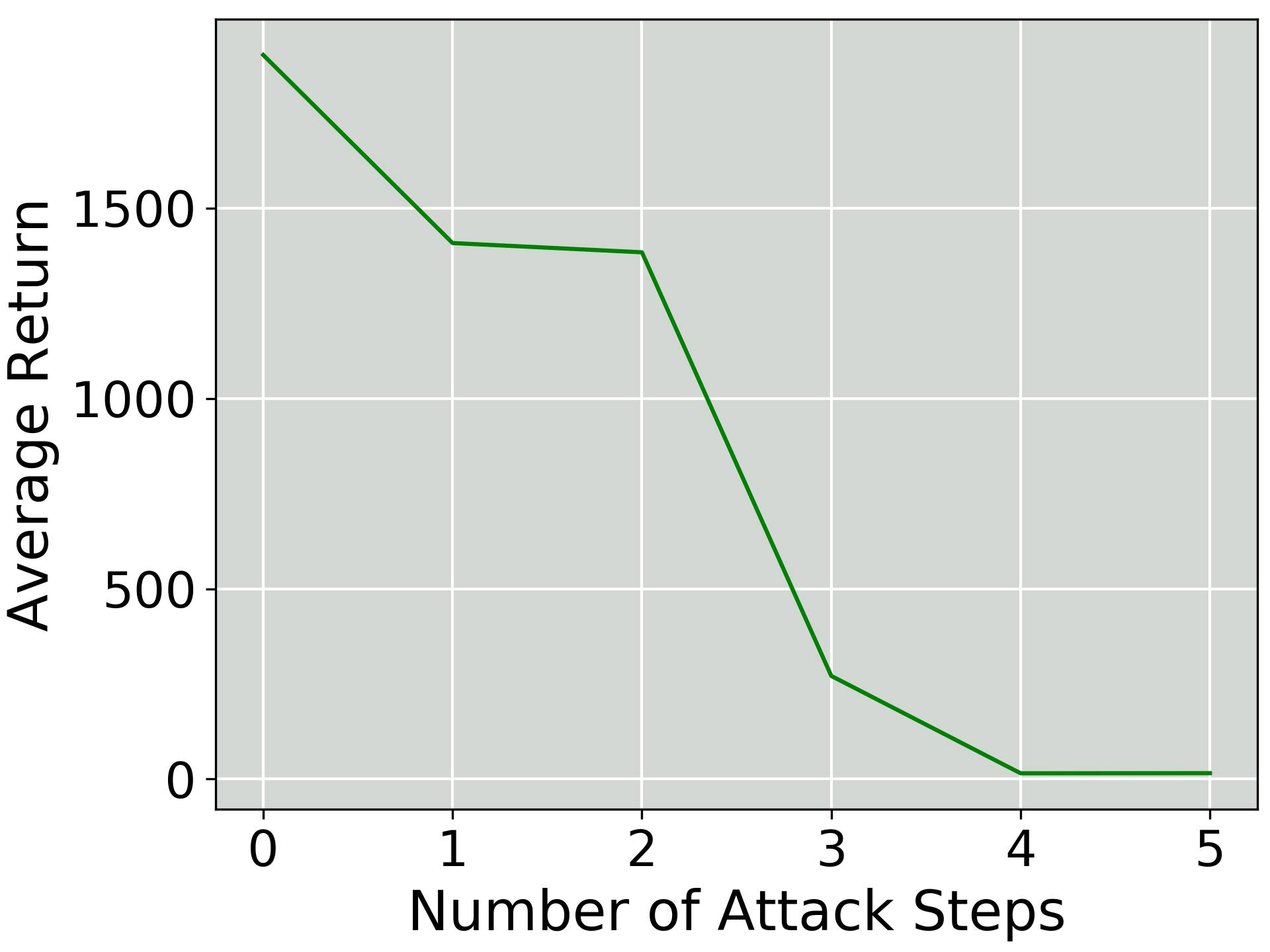}
    \caption{Walker2d}
    \label{anta-walker}
  \end{subfigure}

  \caption{Experimental results of the antagonist attack.}
  \label{anta}
\end{figure}

For each environment, we train multiple versions of antagonists by selecting various numbers 
of attack steps ($N = 1,...,5$). The results are shown in Fig. \ref{anta}. 

\bheading{Atari games.}
For both Pong and Breakout, Antagonist Attack can break down the agent using 3 steps in 
one life cycle. This result is much better compared to strategically-timed attack (6 
steps for Pong and 5 steps for Breakout). Antagonist Attack cannot beat Critical 
Point Attack (2 steps), as it needs to explore the optimal strategy in the state space 
without domain knowledge.

\bheading{Mujoco.}
To reduce the rewards significantly, the numbers of attack steps required for each 
environment are 2 for InvertedPendulum (Fig. \ref{anta-inverted}), 1 for Hopper (Fig.
\ref{anta-hopper}), 5 for HalfCheetah (Fig. \ref{anta-half}) and 3 for Walker2d (Fig. \ref{anta-walker}). Considering that each step takes 
about 4ms in Mujoco environment, these results show that our trained antagonist can break 
down the agent within 20ms.

\subsection{Summary}
Our evaluation results answer the questions proposed at the beginning of this section. 
For \textbf{RQ1}, we show that both of the two attacks can effectively compromise different
DRL tasks (Atari games, TORCS, Mojuco) and algorithms (A3C, DDPG, PPO). We believe they
can be applied to other DRL scenarios and environments. 

For \textbf{RQ2}, we show that our techniques can beat past work in several aspects. (1)
Our attacks require much fewer time steps to perform a successful attack. For Atari games,
Critical Point Attack needs 1 or 2 steps, and Antagonist Attack needs 3 steps, while 
strategically-timed attack \cite{conf/ijcai/LinHLSLS17} requires 5 or 6 steps to achieve 
similar effects. For TORCS, Critical Point Attack needs 1 step, while strategically-time attack requires about 80 steps. This is a significant improvement for the 
attack efficiency. (2) The damage metric adopted in our techniques is more accurate and 
explicit (Fig.~\ref{pm-tac-pong}, \ref{pm-tac-breakout}, \ref{torcs-both}). This enables the
adversary to always make the optimal selection for the critical attack moments. (3) Our 
techniques are generic for different tasks and algorithms, while strategically-time attack
can only work for specific algorithms (A3C and DQN) and specific tasks (discrete action 
space). It cannot attack the agent for a task of continuous action space, e.g., DDPG 
algorithm.

\section{Conclusion}

In this paper, we propose two novel techniques to realize stealthy and efficient adversarial attacks
against Deep Reinforcement Learning systems. These approaches can help the adversary identify the most
critical moments to add perturbations, incurring the most severe damage. We evaluate our approaches
in different types of DRL environments and demonstrate that they need much fewer attack steps
to achieve the same effects compared to past work. 
As future work, we plan to train the antagonist policy with multiple safety goals and enhance the robustness of DRL algorithms.

\paragraph{Acknowledgment} This research was supported by Singapore National Research Foundation, under its National Cybersecurity
 R\&D Program No. NRF2018NCR-NCR005-0001,
 National Satellite of Excellence in Trustworthy Software System No. NRF2018NCR-NSOE003-0001, NTU research grant NGF-2019-06-024, JSPS KAKENHI Grant No.19K24348, 19H04086, and Qdaijump Research Program No.01277.

\balance 

\bibliographystyle{aaai}
\bibliography{references}

\begin{thebibliography}{}

\bibitem[\protect\citeauthoryear{Alzantot, Balaji, and
  Srivastava}{2018}]{alzantot2018did}
Alzantot, M.; Balaji, B.; and Srivastava, M.
\newblock 2018.
\newblock Did you hear that? adversarial examples against automatic speech
  recognition.
\newblock {\em arXiv preprint arXiv:1801.00554}.

\bibitem[\protect\citeauthoryear{Alzantot \bgroup et al\mbox.\egroup
  }{2018}]{alzantot2018generating}
Alzantot, M.; Sharma, Y.; Elgohary, A.; Ho, B.-J.; Srivastava, M.; and Chang,
  K.-W.
\newblock 2018.
\newblock Generating natural language adversarial examples.
\newblock {\em arXiv preprint arXiv:1804.07998}.

\bibitem[\protect\citeauthoryear{Behzadan and
  Munir}{2017a}]{behzadan2017vulnerability}
Behzadan, V., and Munir, A.
\newblock 2017a.
\newblock Vulnerability of deep reinforcement learning to policy induction
  attacks.
\newblock In {\em International Conference on Machine Learning and Data Mining
  in Pattern Recognition},  262--275.
\newblock Springer.

\bibitem[\protect\citeauthoryear{Behzadan and
  Munir}{2017b}]{behzadan2017whatever}
Behzadan, V., and Munir, A.
\newblock 2017b.
\newblock Whatever does not kill deep reinforcement learning, makes it
  stronger.
\newblock {\em arXiv preprint arXiv:1712.09344}.

\bibitem[\protect\citeauthoryear{Carlini and Wagner}{2017}]{carlini2017towards}
Carlini, N., and Wagner, D.
\newblock 2017.
\newblock Towards evaluating the robustness of neural networks.
\newblock In {\em 2017 IEEE Symposium on Security and Privacy (SP)},  39--57.
\newblock IEEE.

\bibitem[\protect\citeauthoryear{Carlini and Wagner}{2018}]{carlini2018audio}
Carlini, N., and Wagner, D.
\newblock 2018.
\newblock Audio adversarial examples: Targeted attacks on speech-to-text.
\newblock In {\em 2018 IEEE Security and Privacy Workshops (SPW)},  1--7.
\newblock IEEE.

\bibitem[\protect\citeauthoryear{Du \bgroup et al\mbox.\egroup
  }{2019}]{du2019deepstellar}
Du, X.; Xie, X.; Li, Y.; Ma, L.; Liu, Y.; and Zhao, J.
\newblock 2019.
\newblock Deepstellar: Model-based quantitative analysis of stateful deep
  learning systems.
\newblock In {\em Proceedings of the 2019 27th ACM Joint Meeting on European
  Software Engineering Conference and Symposium on the Foundations of Software
  Engineering},  477--487.
\newblock ACM.

\bibitem[\protect\citeauthoryear{Eykholt \bgroup et al\mbox.\egroup
  }{2017}]{eykholt2017robust}
Eykholt, K.; Evtimov, I.; Fernandes, E.; Li, B.; Rahmati, A.; Xiao, C.;
  Prakash, A.; Kohno, T.; and Song, D.
\newblock 2017.
\newblock Robust physical-world attacks on deep learning models.
\newblock {\em arXiv preprint arXiv:1707.08945}.

\bibitem[\protect\citeauthoryear{Goodfellow, Shlens, and Szegedy}{2015}]{fgsm}
Goodfellow, I.; Shlens, J.; and Szegedy, C.
\newblock 2015.
\newblock Explaining and harnessing adversarial examples.
\newblock In {\em International Conference on Learning Representations}.

\bibitem[\protect\citeauthoryear{Guo \bgroup et al\mbox.\egroup
  }{2019}]{guo2019empirical}
Guo, Q.; Chen, S.; Xie, X.; Ma, L.; Hu, Q.; Liu, H.; Liu, Y.; Zhao, J.; and Li,
  X.
\newblock 2019.
\newblock An empirical study towards characterizing deep learning development
  and deployment across different frameworks and platforms.
\newblock {\em Proc. 34th IEEE/ACM Conference on Automated Software Engineering
  (ASE 2019)}.

\bibitem[\protect\citeauthoryear{Huang \bgroup et al\mbox.\egroup
  }{2017}]{huang2017adversarial}
Huang, S.; Papernot, N.; Goodfellow, I.; Duan, Y.; and Abbeel, P.
\newblock 2017.
\newblock Adversarial attacks on neural network policies.
\newblock {\em arXiv preprint arXiv:1702.02284}.

\bibitem[\protect\citeauthoryear{Hussenot, Geist, and
  Pietquin}{2019}]{hussenot2019targeted}
Hussenot, L.; Geist, M.; and Pietquin, O.
\newblock 2019.
\newblock Targeted attacks on deep reinforcement learning agents through
  adversarial observations.
\newblock {\em arXiv preprint arXiv:1905.12282}.

\bibitem[\protect\citeauthoryear{Kos and Song}{2017}]{kos2017delving}
Kos, J., and Song, D.
\newblock 2017.
\newblock Delving into adversarial attacks on deep policies.
\newblock {\em arXiv preprint arXiv:1705.06452}.

\bibitem[\protect\citeauthoryear{Kurakin, Goodfellow, and
  Bengio}{2016}]{kurakin2016adversarial}
Kurakin, A.; Goodfellow, I.; and Bengio, S.
\newblock 2016.
\newblock Adversarial examples in the physical world.
\newblock {\em arXiv preprint arXiv:1607.02533}.

\bibitem[\protect\citeauthoryear{Levine \bgroup et al\mbox.\egroup
  }{2016}]{levine2016end}
Levine, S.; Finn, C.; Darrell, T.; and Abbeel, P.
\newblock 2016.
\newblock End-to-end training of deep visuomotor policies.
\newblock {\em The Journal of Machine Learning Research}.

\bibitem[\protect\citeauthoryear{Lillicrap \bgroup et al\mbox.\egroup
  }{2015}]{ddpg}
Lillicrap, T.~P.; Hunt, J.~J.; Pritzel, A.; Heess, N.; and Wierstra, D.
\newblock 2015.
\newblock Continuous control with deep reinforcement learning.
\newblock {\em Computer Science} 8(6):A187.

\bibitem[\protect\citeauthoryear{Lin \bgroup et al\mbox.\egroup
  }{2017}]{conf/ijcai/LinHLSLS17}
Lin, Y.; Hong, Z.; Liao, Y.; Shih, M.; Liu, M.; and Sun, M.
\newblock 2017.
\newblock Tactics of adversarial attack on deep reinforcement learning agents.
\newblock In {\em Proceedings of the Twenty-Sixth International Joint
  Conference on Artificial Intelligence, {IJCAI} 2017, Melbourne, Australia,
  August 19-25, 2017},  3756--3762.

\bibitem[\protect\citeauthoryear{Ma \bgroup et al\mbox.\egroup
  }{2018}]{ma2018deepgauge}
Ma, L.; Juefei-Xu, F.; Zhang, F.; Sun, J.; Xue, M.; Li, B.; Chen, C.; Su, T.;
  Li, L.; Liu, Y.; et~al.
\newblock 2018.
\newblock Deepgauge: Multi-granularity testing criteria for deep learning
  systems.
\newblock In {\em Proceedings of the 33rd ACM/IEEE International Conference on
  Automated Software Engineering},  120--131.
\newblock ACM.

\bibitem[\protect\citeauthoryear{Mnih \bgroup et al\mbox.\egroup
  }{2015}]{Mnih2015HumanlevelCT}
Mnih, V.; Kavukcuoglu, K.; Silver, D.; Rusu, A.~A.; Veness, J.; Bellemare,
  M.~G.; Graves, A.; Riedmiller, M.~A.; Fidjeland, A.; Ostrovski, G.; Petersen,
  S.; Beattie, C.; Sadik, A.; Antonoglou, I.; King, H.; Kumaran, D.; Wierstra,
  D.; Legg, S.; and Hassabis, D.
\newblock 2015.
\newblock Human-level control through deep reinforcement learning.
\newblock {\em Nature} 518:529--533.

\bibitem[\protect\citeauthoryear{Mnih \bgroup et al\mbox.\egroup
  }{2016}]{mnih2016asynchronous}
Mnih, V.; Badia, A.~P.; Mirza, M.; Graves, A.; Lillicrap, T.; Harley, T.;
  Silver, D.; and Kavukcuoglu, K.
\newblock 2016.
\newblock Asynchronous methods for deep reinforcement learning.
\newblock In {\em International conference on machine learning},  1928--1937.

\bibitem[\protect\citeauthoryear{Oh \bgroup et al\mbox.\egroup
  }{2015}]{oh2015action}
Oh, J.; Guo, X.; Lee, H.; Lewis, R.~L.; and Singh, S.
\newblock 2015.
\newblock Action-conditional video prediction using deep networks in atari
  games.
\newblock In {\em Advances in neural information processing systems},
  2863--2871.

\bibitem[\protect\citeauthoryear{Papernot \bgroup et al\mbox.\egroup
  }{2016}]{papernot2016limitations}
Papernot, N.; McDaniel, P.; Jha, S.; Fredrikson, M.; Celik, Z.~B.; and Swami,
  A.
\newblock 2016.
\newblock The limitations of deep learning in adversarial settings.
\newblock In {\em 2016 IEEE European Symposium on Security and Privacy
  (EuroS\&P)},  372--387.
\newblock IEEE.

\bibitem[\protect\citeauthoryear{Pattanaik \bgroup et al\mbox.\egroup
  }{2018}]{pattanaik2018robust}
Pattanaik, A.; Tang, Z.; Liu, S.; Bommannan, G.; and Chowdhary, G.
\newblock 2018.
\newblock Robust deep reinforcement learning with adversarial attacks.
\newblock In {\em Proceedings of the 17th International Conference on
  Autonomous Agents and MultiAgent Systems},  2040--2042.
\newblock International Foundation for Autonomous Agents and Multiagent
  Systems.

\bibitem[\protect\citeauthoryear{Russo and Proutiere}{2019}]{russo2019optimal}
Russo, A., and Proutiere, A.
\newblock 2019.
\newblock Optimal attacks on reinforcement learning policies.
\newblock {\em arXiv preprint arXiv:1907.13548}.

\bibitem[\protect\citeauthoryear{Sallab \bgroup et al\mbox.\egroup
  }{2017}]{sallab2017deepcar}
Sallab, A.~E.; Abdou, M.; Perot, E.; and Yogamani, S.
\newblock 2017.
\newblock Deep reinforcement learning framework for autonomous driving.
\newblock {\em Electronic Imaging} 2017(19):70--76.

\bibitem[\protect\citeauthoryear{Schulman \bgroup et al\mbox.\egroup
  }{2017}]{ppo}
Schulman, J.; Wolski, F.; Dhariwal, P.; Radford, A.; and Klimov, O.
\newblock 2017.
\newblock Proximal policy optimization algorithms.
\newblock {\em arXiv preprint arXiv:1707.06347}.

\bibitem[\protect\citeauthoryear{Sharif \bgroup et al\mbox.\egroup
  }{2016}]{sharif2016accessorize}
Sharif, M.; Bhagavatula, S.; Bauer, L.; and Reiter, M.~K.
\newblock 2016.
\newblock Accessorize to a crime: Real and stealthy attacks on state-of-the-art
  face recognition.
\newblock In {\em Proceedings of the 2016 ACM SIGSAC Conference on Computer and
  Communications Security},  1528--1540.
\newblock ACM.

\bibitem[\protect\citeauthoryear{Silver \bgroup et al\mbox.\egroup
  }{2016}]{Silver2016MasteringTG}
Silver, D.; Huang, A.; Maddison, C.~J.; Guez, A.; Sifre, L.; van~den Driessche,
  G.; Schrittwieser, J.; Antonoglou, I.; Panneershelvam, V.; Lanctot, M.;
  Dieleman, S.; Grewe, D.; Nham, J.; Kalchbrenner, N.; Sutskever, I.;
  Lillicrap, T.~P.; Leach, M.; Kavukcuoglu, K.; Graepel, T.; and Hassabis, D.
\newblock 2016.
\newblock Mastering the game of go with deep neural networks and tree search.
\newblock {\em Nature} 529:484--489.

\bibitem[\protect\citeauthoryear{Szegedy \bgroup et al\mbox.\egroup
  }{2013}]{szegedy2013intriguing}
Szegedy, C.; Zaremba, W.; Sutskever, I.; Bruna, J.; Erhan, D.; Goodfellow, I.;
  and Fergus, R.
\newblock 2013.
\newblock Intriguing properties of neural networks.
\newblock {\em arXiv preprint arXiv:1312.6199}.

\bibitem[\protect\citeauthoryear{Szegedy \bgroup et al\mbox.\egroup
  }{2014}]{Szegedy2014IntriguingPO}
Szegedy, C.; Zaremba, W.; Sutskever, I.; Bruna, J.; Erhan, D.; Goodfellow,
  I.~J.; and Fergus, R.
\newblock 2014.
\newblock Intriguing properties of neural networks.
\newblock {\em International Conference on Learning Representations (ICLR)}.

\bibitem[\protect\citeauthoryear{Szegedy \bgroup et al\mbox.\egroup
  }{2015}]{szegedy2015going}
Szegedy, C.; Liu, W.; Jia, Y.; Sermanet, P.; Reed, S.; Anguelov, D.; Erhan, D.;
  Vanhoucke, V.; and Rabinovich, A.
\newblock 2015.
\newblock Going deeper with convolutions.
\newblock In {\em Proceedings of the IEEE conference on computer vision and
  pattern recognition},  1--9.

\bibitem[\protect\citeauthoryear{Tretschk, Oh, and
  Fritz}{2018}]{tretschk2018sequential}
Tretschk, E.; Oh, S.~J.; and Fritz, M.
\newblock 2018.
\newblock Sequential attacks on agents for long-term adversarial goals.
\newblock {\em arXiv preprint arXiv:1805.12487}.

\bibitem[\protect\citeauthoryear{Xie \bgroup et al\mbox.\egroup
  }{2019a}]{xie2019deephunter}
Xie, X.; Ma, L.; Juefei-Xu, F.; Xue, M.; Chen, H.; Liu, Y.; Zhao, J.; Li, B.;
  Yin, J.; and See, S.
\newblock 2019a.
\newblock Deephunter: a coverage-guided fuzz testing framework for deep neural
  networks.
\newblock In {\em Proceedings of the 28th ACM SIGSOFT International Symposium
  on Software Testing and Analysis},  146--157.
\newblock ACM.

\bibitem[\protect\citeauthoryear{Xie \bgroup et al\mbox.\egroup
  }{2019b}]{xie2019diffchaser}
Xie, X.; Ma, L.; Wang, H.; Li, Y.; Liu, Y.; and Li, X.
\newblock 2019b.
\newblock Diffchaser: detecting disagreements for deep neural networks.
\newblock In {\em Proceedings of the 28th International Joint Conference on
  Artificial Intelligence},  5772--5778.
\newblock AAAI Press.

\bibitem[\protect\citeauthoryear{Zheng \bgroup et al\mbox.\egroup
  }{2018a}]{zheng2018deep}
Zheng, Y.; Meng, Z.; Hao, J.; Zhang, Z.; Yang, T.; and Fan, C.
\newblock 2018a.
\newblock A deep bayesian policy reuse approach against non-stationary agents.
\newblock In {\em Advances in Neural Information Processing Systems},
  954--964.

\bibitem[\protect\citeauthoryear{Zheng \bgroup et al\mbox.\egroup
  }{2018b}]{zheng2018weighted}
Zheng, Y.; Meng, Z.; Hao, J.; and Zhang, Z.
\newblock 2018b.
\newblock Weighted double deep multiagent reinforcement learning in stochastic
  cooperative environments.
\newblock In {\em Pacific Rim International Conference on Artificial
  Intelligence},  421--429.
\newblock Springer.

\bibitem[\protect\citeauthoryear{Zheng \bgroup et al\mbox.\egroup
  }{2019a}]{zheng2019diverse}
Zheng, Y.; Shen, R.; Hao, J.; Chen, Y.; and Fan, C.
\newblock 2019a.
\newblock Diverse behavior is what game ai needs: Generating varied human-like
  playing styles using evolutionary multi-objective deep reinforcement
  learning.
\newblock {\em arXiv preprint arXiv:1910.09022}.

\bibitem[\protect\citeauthoryear{Zheng \bgroup et al\mbox.\egroup
  }{2019b}]{zheng2019wuji}
Zheng, Y.; Xie, X.; Su, T.; Ma, L.; Hao, J.; Meng, Z.; Liu, Y.; Shen, R.; Chen,
  Y.; and Fan, C.
\newblock 2019b.
\newblock Wuji: Automatic online combat game testing using evolutionary deep
  reinforcement learning.
\newblock In {\em Proceedings of the 34th ACM/IEEE International Conference on
  Automated Software Engineering}.

\end{thebibliography}

\end{document}